\documentclass{article}
\usepackage[utf8]{inputenc}
\usepackage[english]{babel}
\usepackage{comment}
\usepackage{authblk}
\usepackage{graphicx}
\graphicspath{{figures/}{../figures/}}
\usepackage{blindtext}
\usepackage{subfiles}
\usepackage{amsmath}
\usepackage{siunitx}
\usepackage{booktabs} 
\usepackage{hyperref}
\usepackage{float}
\usepackage{textcomp}
\usepackage[sort,comma,authoryear,round]{natbib}
\usepackage{tabularx}
\usepackage{chngcntr}
\usepackage{fancyhdr}
\usepackage[font=small,skip=0pt]{caption}
\usepackage[letterpaper, left=1.3in, right=1.3in, bottom=1.25in, top=1.25in]{geometry}
\usepackage{lineno}

\title{Evaluating deep transfer learning for\\whole-brain cognitive decoding}

\author[1,2,3,4]{Armin W. Thomas}
\author[2,5]{Ulman Lindenberger}
\author[6,7]{Wojciech Samek}
\author[1,7,8,9]{Klaus-Robert M{\"u}ller}

\affil[1]{Machine Learning Group, Dept.\ of Computer Science and Electrical Engineering, Technische Universit{\"a}t Berlin, Berlin, Germany}
\affil[2]{Center for Lifespan Psychology, Max Planck Institute for Human Development, Berlin, Germany}
\affil[3]{ Stanford Data Science, Stanford University, Stanford, CA, USA}
\affil[4]{Dept.\ of Psychology, Stanford University, Stanford, CA, USA}
\affil[5]{Max Planck UCL Centre for Computational Psychiatry and Ageing Research, Berlin, Germany, and London, UK}
\affil[6]{Dept.\ of Artificial Intelligence, Fraunhofer Heinrich Hertz Institute, Berlin, Germany}
\affil[7]{BIFOLD – Berlin Institute for the Foundations of Learning and Data, Berlin, Germany}
\affil[8]{Dept. of Artificial Intelligence, Korea University, Seoul, South Korea}
\affil[9]{Max Planck Institute for Informatics,  Saarbr{\"u}cken, Germany}

\date{November 2021}

\begin{document}

\maketitle

\begin{abstract}
    Research in many fields has shown that transfer learning (TL) is well-suited to improve the performance of deep learning (DL) models in datasets with small numbers of samples.
    This empirical success has triggered interest in the application of TL to cognitive decoding analyses with functional neuroimaging data.
    Here, we systematically evaluate TL for the application of DL models to the decoding of cognitive states (e.g., viewing images of faces or houses) from whole-brain functional Magnetic Resonance Imaging (fMRI) data.
    We first pre-train two DL architectures on a large, public fMRI dataset and subsequently evaluate their performance in an independent experimental task and a fully independent dataset.
    The pre-trained models consistently achieve higher decoding accuracies and generally require less training time and data than model variants that were not pre-trained, clearly underlining the benefits of pre-training.
    We demonstrate that these benefits arise from the ability of the pre-trained models to reuse many of their learned features when training with new data, providing deeper insights into the mechanisms giving rise to the benefits of pre-training.
    Yet, we also surface nuanced challenges for whole-brain cognitive decoding with DL models when interpreting the decoding decisions of the pre-trained models, as these have learned to utilize the fMRI data in unforeseen and counterintuitive ways to identify individual cognitive states.
    
    \medskip
    
\end{abstract}

\newpage
\section{Introduction}
One of the key challenges for the analysis of functional Magnetic Resonance Imaging (fMRI) data is the high dimensionality and low sample size of conventional fMRI datasets.
These datasets typically contain several hundred thousand dimensions (or voxels) for each fMRI volume, while containing only up to a few hundred volumes for each of up to a hundred individuals.
To prevent overfitting in these high-dimensional and low-sample size settings, allowing for generalizable statistical inference, conventional approaches for the analysis of fMRI data often include restricting assumptions, by analyzing the data of individual voxels or groups of voxels independent of one another, using simple linear mappings between the cognitive states and brain activity, or by solely focusing on the group level \citep[e.g.,][]{friston_statistical_1994,kriegeskorte_information-based_2006}.
While these restrictions are well justified for many research questions (e.g., to test whether a specific region-of-interest exhibits more activity in one experimental condition than in another), they also limit the ability of these approaches to capture the temporal and spatial variability of the mapping between a cognitive state (e.g., deciding whether to accept or reject a risky gamble) and the underlying whole-brain brain activity within and between individuals.

Research in many other fields has shown that deep learning \citep[DL;][]{lecun_deep_2015,goodfellow_deep_2016,schmidhuber_deep_2015} methods are generally well-suited to capture complex nonlinear mappings between a target signal (e.g., a cognitive state) and highly variable patterns in the data.
For this reason, researchers have started exploring their application to cognitive decoding \citep[e.g.,][]{thomas_analyzing_2019,abrol_deep_2021,wang_decoding_2020,zhang_functional_2021,riaz_deepfmri_2020,suk_state-space_2016,schirrmeister_deep_2017,mahmood_whole_2020}, by training these models to identify (or decode) a set of cognitive states from fMRI data.
To subsequently identify an association between the decoded cognitive states and measured brain activity, researchers have turned towards research on explainable artificial intelligence \citep[XAI;][]{samek_explaining_2021,montavon_methods_2018,samek_explainable_2017}.
First empirical work has demonstrated that this combination of DL and XAI is well-suited for the analysis of fMRI data, by accurately decoding cognitive states and identifying biologically plausible associations between the decoded cognitive states and brain activity through the interpretation of a model's decoding decisions \citep{thomas_analyzing_2019,zhang_functional_2021,mcclure_evaluating_2020,koyamada_deep_2015,dinsdale_learning_2021,nguyen_attend_2020}.
Yet, many of these studies have also demonstrated that the advantages of DL methods over conventional approaches for the analysis of fMRI data are generally limited to sufficiently large training datasets \citep{thomas_analyzing_2019,abrol_deep_2021,schulz_different_2020}.

The problem of limited data is not new to the field of machine learning (ML), where researchers have discovered that transfer learning can generally improve the performance of DL models in small datasets.
The goal of transfer learning is to leverage the knowledge about the mapping between input data and a target signal that can be learned from one dataset to subsequently improve the learning of a similar mapping in another dataset of a related domain.
Transfer learning has been particularly successful in computer vision and natural language processing, where large, publicly available datasets exist \citep{deng_imagenet_2009,bowman_large_2015,rajaraman_pre-trained_2018}.
These datasets are used to first \emph{pre-train} DL models (e.g., to identify the objects in an image), before \emph{fine-tuning} them on smaller datasets of a related domain \citep[e.g., to detect breast cancer in medical imaging;][]{khan_novel_2019}.
When compared to models that are trained from scratch, pre-trained models generally exhibit faster learning and achieve higher predictive accuracies, while also requiring less training data \citep{yosinski_how_2014,raffel_exploring_2020,kolesnikov_big_2020,chen_big_2020,bengio_greedy_2006,erhan_why_2010}.

Over recent years, neuroimaging research has experienced a strong increase in the availability of public datasets.
These datasets are provided by large neuroimaging initiatives \citep[e.g.,][]{van_essen_wu-minn_2013,casey_adolescent_2018,sudlow_uk_2015,poldrack_phenome-wide_2016} and individual researchers \citep[e.g.,][]{markiewicz_openneuro_2021}.
Due to the availability of these datasets, researchers have started exploring whether transfer learning can be beneficial for cognitive decoding analysis with fMRI data.
This work generally indicates that the pre-trained models similarly achieve higher decoding accuracies, and require less training time and data, when compared to models that are trained from scratch \citep{zhang_functional_2021,koyamada_deep_2015,svanera_transfer_2019,mahmood_learnt_2019,thomas_deep_2019,oh_classification_2019,deepak_brain_2019,mensch_extracting_2021}. 
Yet, this work has often been limited by not evaluating the performance of the pre-trained models in fully independent datasets \citep[e.g., by pre-training on some part of a large dataset and evaluating on another part of the same dataset;][]{zhang_functional_2021,koyamada_deep_2015,thomas_deep_2019}, by using very simple decoding tasks \citep[e.g., by only identifying the dataset or experimental task underlying the fMRI data;][]{koyamada_deep_2015} and experiments \citep[e.g., requiring the experiment to use visual image stimuli;][]{svanera_transfer_2019}, and by applying strong feature preprocessing to the fMRI data \citep[e.g., independent component, general linear model, and connectivity analyses;][]{mahmood_learnt_2019,mensch_extracting_2021,schulz_different_2020,mahmood_whole_2020}.
Further, many studies have not made their pre-trained DL models available to the public, making it difficult to evaluate the performance of these models in other datasets.

Here, we systematically evaluate deep transfer learning for whole-brain cognitive decoding, by first pre-training two DL architectures on a large whole-brain fMRI dataset and subsequently comparing their performances on an independent experimental task and a fully independent fMRI dataset.
We also provide detailed insights into the mechanisms giving rise to the advantages of the pre-trained models, bring forward nuanced challenges for whole-brain cognitive decoding with DL models, and make our pre-trained models available to the public. 

Specifically, we train two distinct DL architectures on task-fMRI data from the Human Connectome Project \citep[HCP;][]{van_essen_wu-minn_2013}, spanning six experimental tasks and 16 distinct cognitive states.
We demonstrate that the pre-trained DL models learn quicker, achieve higher decoding accuracies, and require less training data than model variants that were not pre-trained, when applied to the fMRI data of the left-out seventh HCP experimental task.
We further evaluate the performance of the pre-trained models in a fully independent fMRI dataset that is not part of the HCP.
Again, the pre-trained models outperform model variants that were not pre-trained by achieving overall higher decoding accuracies and generally learning quicker.
To better understand the mechanisms giving rise to the advantages of the pre-trained models, we perform an analysis of their learned hidden representations.
This analysis reveals that these advantages generally arise from the ability of the pre-trained models to reuse many of their learned features when training with new data.
To also understand the models' learned mappings between brain activity and cognitive state, we next interpret their cognitive decoding decisions for the fMRI data of the different HCP experimental tasks.
We find that the pre-trained models generally associate a similar set of brain regions with the cognitive states of these tasks as a standard general linear model \citep[GLM;][]{holmes_generalisability_1998} analysis of the same data. 
Interestingly, the models identify individual cognitive states by combining activity from brain regions that are generally more active in these states, when compared to other states, with the activity of brain regions that are generally less active in these states.
While this represents a biologically plausible solution for the underlying decoding task, the resulting brain maps are, at first sight, surprising, as the models assign high relevance to brain regions that the GLM analysis associates negatively with the decoded cognitive states.
Importantly, this pattern only then becomes apparent, when interpreting the models' decoding decisions in light of the results of the GLM analysis.

In sum, our results highlight that DL models are capable of learning versatile representations of large fMRI datasets, which generalize well to new data and thus allow for successful transfer learning.
Yet, these benefits come at a cost, as the underlying learned mappings between brain activity and cognitive states can be complex, unforeseen, and counterintuitive, requiring careful additional analyses to be understood well.

\section{Results}
\subsection{The DeepLight framework}
\label{sec:results:deeplight-introduction}

\begin{figure}[!t]
\begin{center}
\includegraphics[width=\linewidth]{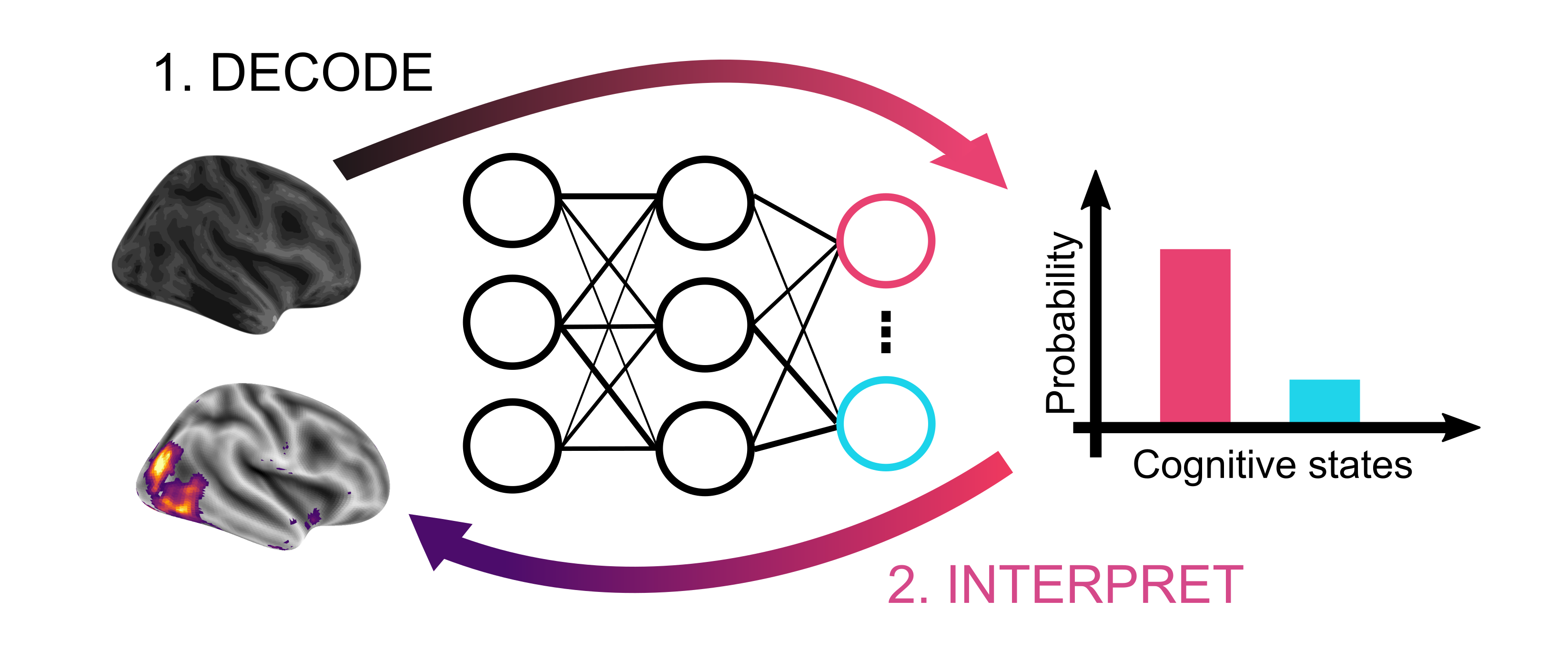}
\end{center}
\caption{The DeepLight framework.
First, DeepLight trains a DL model to accurately decode a set of cognitive states (e.g., viewing the image of a house or face) from single whole-brain fMRI volumes.
Subsequently, DeepLight relates the decoded cognitive states and brain activity by interpreting the decoding decisions of the DL model with methods from explainable artificial intelligence, thereby quantifying the contribution of individual input voxel activities to the resulting decoding decision.}
\label{fig:deeplight-framework}
\end{figure}

\begin{figure}[!t]
\begin{center}
\includegraphics[width=\linewidth]{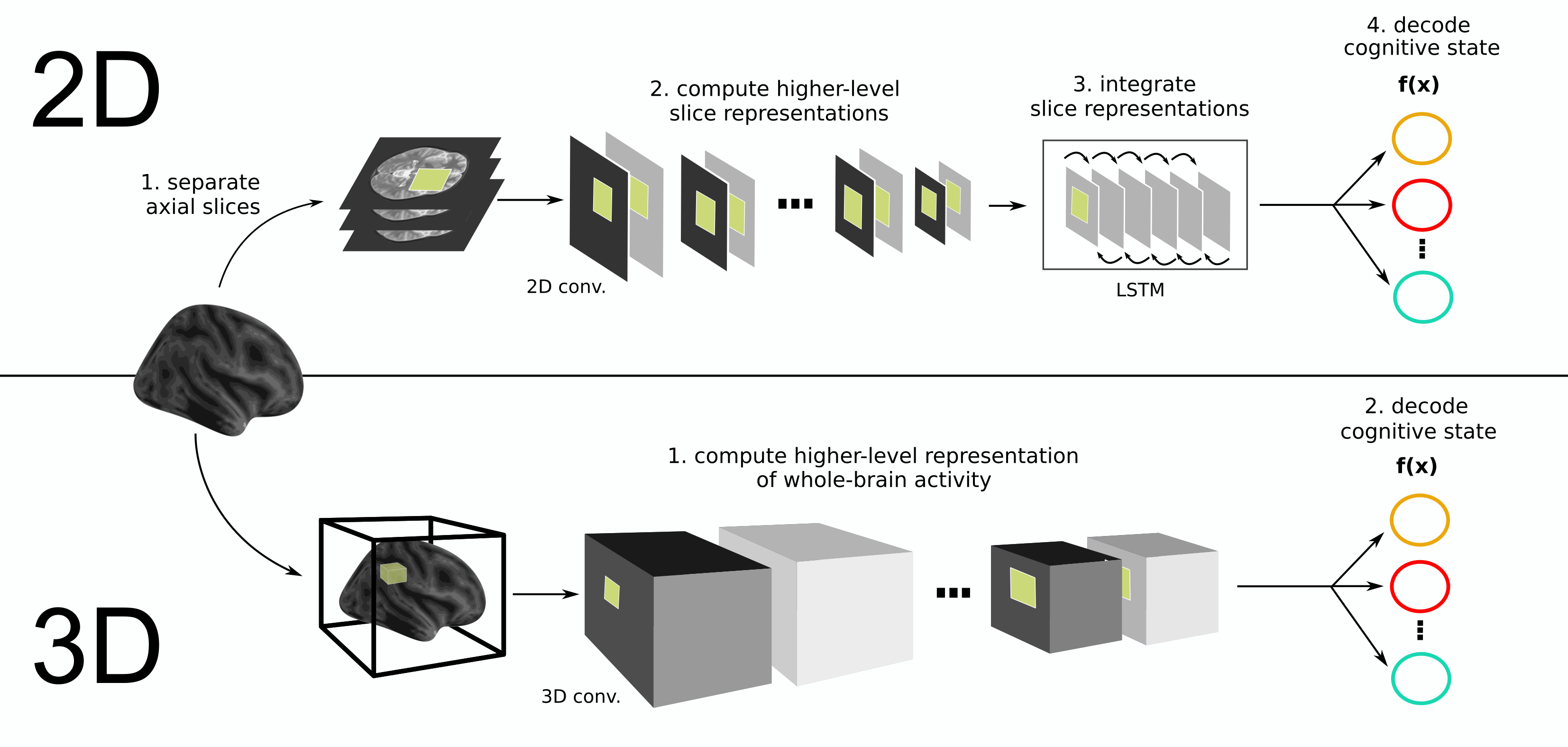}
\end{center}
\caption{Tested DeepLight architectures.
\emph{2D-DeepLight}: A whole-brain fMRI volume is sliced into a sequence of axial images.
These images are passed to a DL model, consisting of a 2D-convolutional feature extractor as well as an LSTM and output layer.
First, the 2D-convolutional feature extractor reduces the dimensionality of the axial brain slices through a sequence of 2D-convolution layers.
The resulting sequence of higher-level slice representations is fed to a bi-directional LSTM, modeling the spatial dependencies of brain activity within and across brain slices.
Lastly, 2D-DeepLight outputs a decoding decision about the cognitive state underlying the fMRI volume, through a softmax output layer with one output unit per cognitive state in the data.
\emph{3D-DeepLight}: A whole-brain fMRI volume is passed to a 3D-convolutional feature extractor, consisting of a sequence of multiple 3D-convolution layers.
The 3D-convolutional feature extractor projects the fMRI data into a higher-level, but lower dimensional, representation of whole-brain activity.
To make a decoding decision, 3D-DeepLight utilizes an output layer, which is composed of a 1D-convolution and global average pooling layer as well as a softmax activation function.
The 1D-convolution layer maps the higher-level representation of whole-brain activity of the 3D-convolutional feature extractor to one representation for each cognitive state in the data.
The global average pooling layer and softmax function reduce these to a decoding decision.}
\label{fig:deeplight-architectures}
\end{figure}

To evaluate the benefits of deep transfer learning for whole-brain cognitive decoding, we utilize the DeepLight framework \citep{thomas_analyzing_2019,thomas_deep_2019}.
DeepLight is defined by two central components (see Fig.\ \ref{fig:deeplight-framework}):
It first trains a DL model to decode a set of cognitive states from a single whole-brain fMRI volumes (e.g., single TRs) and subsequently relates the decoded cognitive states and brain activity by interpreting the decoding decisions with methods from XAI.
Here, we use the layer-wise relevance propagation \citep[LRP; ][]{bach_pixel-wise_2015,montavon_explaining_2017} technique, which decomposes individual decoding decisions of a DL model into the contributions of the activity of each input voxel to the decisions.

Note that DeepLight is not restricted to any specific DL model architecture.
In this work, we explore and compare the performance of two distinct DL model architectures (here abbreviated as "2D-DeepLight" and "3D-DeepLight"; see Fig.\ \ref{fig:deeplight-architectures} and section \ref{sec:methods:deeplight-architectures}), which are based on recent empirical work in computer vision \citep{donahue_long-term_2017,marban_recurrent_2019,tran_learning_2015}.

\subsection{DeepLight accurately decodes cognitive states of the pre-training data}
\label{sec:results:pre-training}

\begin{figure}[!t]
\begin{center}
\includegraphics[width=\linewidth]{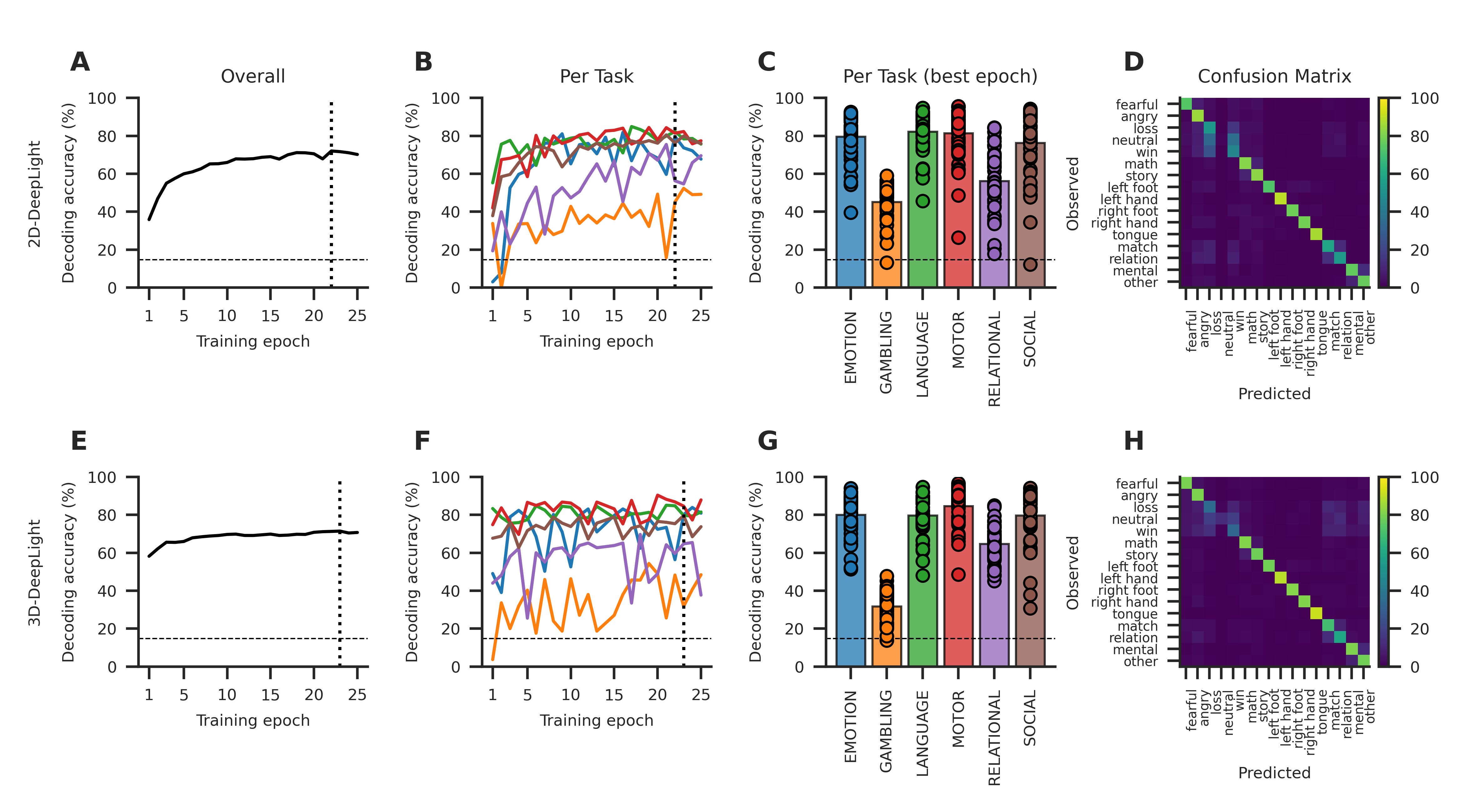}
\end{center}
\caption{Decoding accuracies of the 2D- (A-D) and 3D-DeepLight (E-H) architectures during pre-training. A-B, E-F:
The 2D- (A-B) and 3D-DeepLight (E-F) architectures both learned to accurately decode the cognitive states from the validation data of the pre-training dataset.
Dotted vertical lines indicate the training epoch with the maximum overall decoding accuracy in the validation data.
C, G: Both architectures generally perform well at decoding the cognitive states of each task, at the pre-training epoch with the highest overall decoding accuracy in the validation dataset (dotted vertical lines in A-B and E-F).
Bar heights indicate overall decoding accuracies, with scatter points indicating individual subjects.
Dashed horizontal lines in A-C and E-G indicate chance level ($14.74\%$).
D, H: Confusion matrices for the 2D- (D) and 3D- (H) DeepLight architectures at the pre-training epoch with the highest overall decoding accuracy in the validation dataset.
Brighter yellow colors indicate fewer errors.}
\label{fig:results:pre-training-performance}
\end{figure}

We first pre-trained a variant of each DeepLight architecture (2D and 3D; see Fig. \ref{fig:deeplight-architectures} and section \ref{sec:methods:deeplight-architectures}) to identify the cognitive states underlying the individual TRs of a large whole-brain fMRI dataset (for details on the training procedures, see section \ref{sec:methods:deeplight-training}), spanning 450 participants in six of the seven HCP experimental tasks (all tasks except for the working memory task; for an overview of the individual experimental tasks, see section \ref{sec:methods:data:hcp} and Appendix \ref{sec:appendix:methods:HCP-tasks}).
Importantly, all TRs of an experimental block were included in our analyses, except for the first TR of each block, which we excluded.

To evaluate the ability of the 2D- and 3D-DeepLight architectures to identify the cognitive states from individual fMRI volumes (i.e., TRs), we further divided the data within each pre-training task into distinct training and validation datasets, by designating the fMRI data of 400 randomly selected subjects as training data and the data of the remaining 50 subjects as a validation dataset.

During pre-training, the output layer of both DeepLight architectures contained 16 units, one for each cognitive state of each task in the pre-training dataset (for an overview of the individual cognitive states, see Table \ref{table:methods:data:hcp}).
The DeepLight architectures therefore had no knowledge of the number of experimental tasks in the dataset and the assignment of cognitive states to each task.
Both architectures were solely trained to identify the 16 cognitive states from single TRs.
We trained each DeepLight architecture for a period of 25 epochs of stochastic gradient descent (Fig.\ \ref{fig:results:pre-training-performance} A, E; for details on the training procedures, see section \ref{sec:methods:deeplight-training}).
Each epoch was defined as an iteration over the entire training data of the pre-training dataset. 

Both DeepLight architectures performed well in decoding the cognitive states from the fMRI data of the pre-training dataset (at a chance level decoding accuracy of $14.74\%$; Fig.\ \ref{fig:results:pre-training-performance} A, E):
2D-DeepLight achieved its highest decoding accuracy in the validation data after 22 training epochs ($72.01\%$; Fig.\ \ref{fig:results:pre-training-performance} A), whereas  3D-DeepLight achieved its highest validation decoding accuracy after 23 training epochs ($71.47\%$; Fig.\ \ref{fig:results:pre-training-performance} E).
Note that these performances were statistically not meaningfully different from one another for the 50 subjects in the validation data of each of the six experimental tasks in the pre-training dataset ($t(299)=-0.17, P=0.86$).
Importantly, we used the parameter estimates from the training epochs at which each DeepLight architecture performed best in the validation data (i.e., 22 for 2D-DeepLight and 23 for 3D-DeepLight; Fig.\ \ref{fig:results:pre-training-performance} A, E) for all subsequent analyses involving the pre-trained models. 

Both DeepLight architectures generally performed best at identifying the cognitive states of the "motor" (2D: $81.43\%$, 3D: $84.64\%$; Fig.\ \ref{fig:results:pre-training-performance} B-C, F-G), "language" (2D: $82.30\%$, 3D: $79.74\%$; Fig.\ \ref{fig:results:pre-training-performance} B-C, F-G), "emotion" (2D: $79.66\%$, 3D: $80.16\%$; Fig.\ \ref{fig:results:pre-training-performance} B-C, F-G), and "social" (2D: $76.25\%$, 3D: $79.68\%$; Fig.\ \ref{fig:results:pre-training-performance} B-C, F-G) HCP experimental tasks (see Fig.\ \ref{fig:results:pre-training-performance} B-C, F-G).
They did not perform as well in the "relational"  experimental task (2D: $56.29\%$, 3D: $64.75\%$; Fig.\ \ref{fig:results:pre-training-performance} B-C, F-G) and generally struggled to accurately decode the cognitive states of the "gambling" task (2D: $45.12\%$, 3D: $31.80\%$; Fig.\ \ref{fig:results:pre-training-performance} B-C, F-G).

Interestingly, both architectures exhibited little confusion between the cognitive states of different experimental tasks (with the exception of the gambling task; see Fig.\ \ref{fig:results:pre-training-performance} D, H), indicating that they were able to correctly group the cognitive states of the tasks without receiving any explicit information about the task structure during training.
\subsection{Pre-trained models transfer well to an independent experimental task}
\label{sec:results:fine-tuning}

\begin{figure}[!t]
\begin{center}
\includegraphics[width=\linewidth]{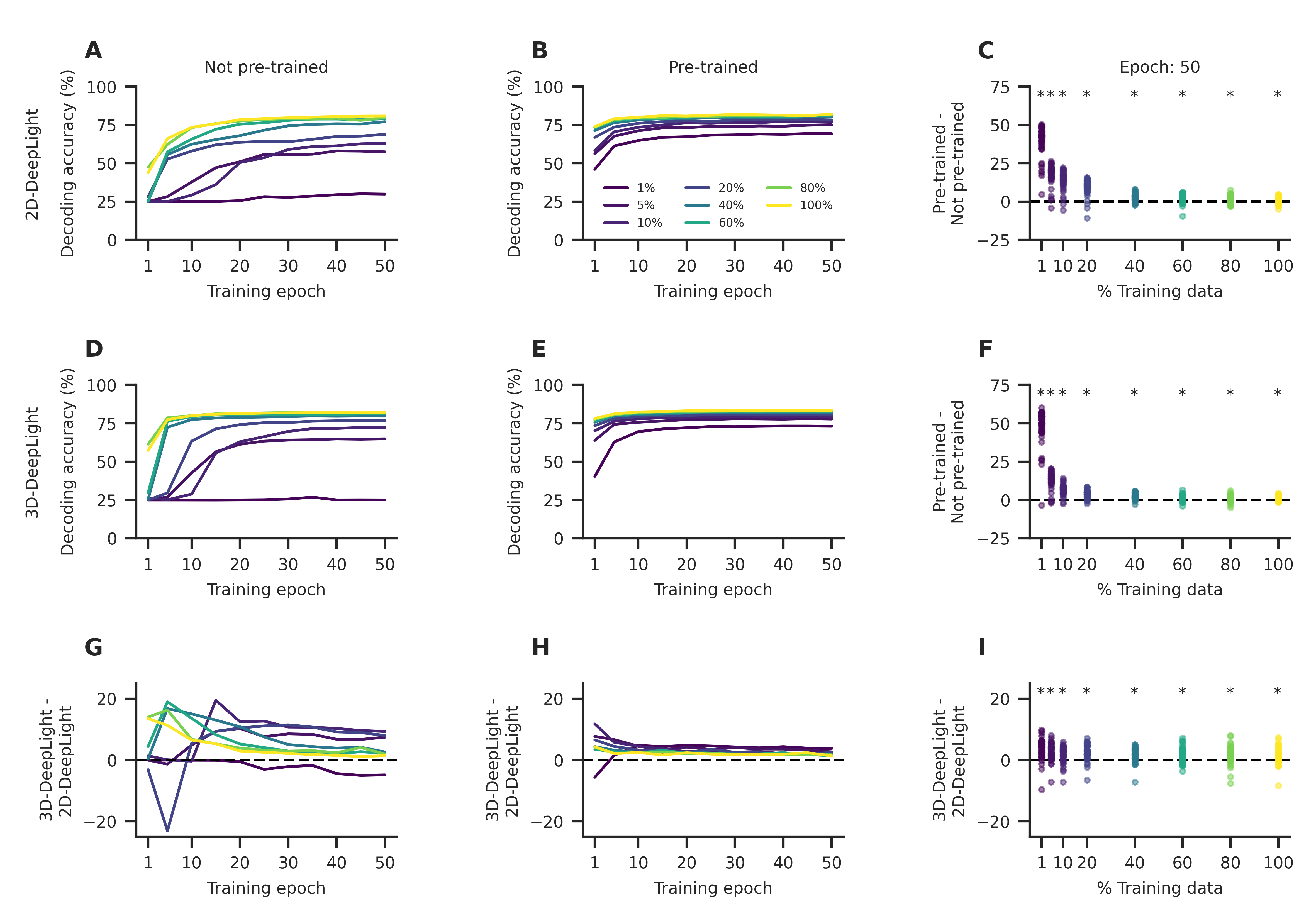}
\end{center}
\caption{Decoding performance of a pre-trained and not pre-trained variant of the 2D- (A-C) and 3D-DeepLight (D-F) architectures in the validation data of the HCP-WM task (N = 50).
We repeatedly trained the pre-trained (A, D) and not pre-trained (B, E) DeepLight variants on the fMRI data of 1\%, 5\%, 10\%, 20\%, 40\%, 60\%, 80\%, and 100\% of the training dataset of the HCP-WM task (N = 400).
After 50 training epochs (C, F), the pre-trained DeepLight variants consistently achieved higher decoding accuracies than their not pre-trained counterparts in the validation data.
G-I: Difference between the decoding accuracies of the 2D- (A-C) and 3D-DeepLight (D-F) architectures.
In general, the 3D-DeepLight architectures achieved higher decoding accuracies than their 2D counterparts.
Lines indicate decoding accuracies in the validation data, while scatter points indicate individual subjects.
Stars indicate that the distribution of subject decoding accuracies is meaningfully different from 0 in a two-sided t-test (Bonferroni-corrected, such that $P \leq 0.05/8$).
Colors indicate the different fractions of the training dataset.}
\label{fig:results:fine-tuning-performance}
\end{figure}

Both pre-trained DeepLight architectures (2D and 3D; see Fig.\ \ref{fig:deeplight-architectures}) accurately decode the cognitive states of the pre-training dataset (see Fig.\ \ref{fig:results:pre-training-performance} C, G).
Next we were therefore interested in evaluating whether they perform better at identifying the four cognitive states of the left-out HCP experimental task (the "working memory" task; see section \ref{sec:methods:data:hcp}) when compared to respective model variants that are trained from scratch.
In the HCP-working memory task (HCP-WM), individuals viewed images of body parts, faces, places, and tools in the fMRI.
The target of the decoding analysis was to identify these four cognitive states from individual TRs.

\subsubsection{Allowing the pre-trained weights to change during fine-tuning is beneficial}
\label{sec:results:fine-tuning:free-vs-frozen}

In a first step of this analysis, we compared the performance of two common fine-tuning approaches, by testing whether a model that freezes the pre-trained weights during fine-tuning \citep[e.g.,][]{rajaraman_pre-trained_2018} performs better than a model variant that continues to train these weights \citep[e.g.,][]{samek_understanding_2017,thomas_deep_2019}.
To do this, we first initialized the weights of two identical variants of each architecture (for an overview of the model architectures, see Fig.\ \ref{fig:deeplight-architectures} and section \ref{sec:methods:deeplight-architectures}) to the weights of the pre-trained models (except for weights of the output layers, which now included four instead of 16 units; see section \ref{sec:methods:deeplight-architectures}).
We then held the pre-trained weights of one variant of each architecture constant during fine-tuning, while we allowed the weights of the respective other variant to change.

After 50 training epochs (for an overview of the training procedures, see section \ref{sec:methods:deeplight-training}), the model variants whose weights were allowed to change clearly outperformed the model variants with frozen weights (see Appendix Fig.\ \ref{fig:appendix:deeplight:transfer:finetuning-approach-comparison}):
While the models with frozen weights achieved final decoding accuracies of $54.73\%$ (2D) and $64.50\%$ (3D) in the validation data of the HCP-WM task, model variants that were allowed to change these weights achieved a final decoding accuracy of $82.10\%$ (2D) and $83.45\%$ (3D).
We therefore allowed the pre-trained models to change their weights during fine-tuning in all further transfer learning analyses.

\subsubsection{Pre-trained models decode more accurately than models trained from scratch}
\label{sec:results:fine-tuning:pre-trained-vs-scratch}

In a second step of this analysis, we then compared the performance of the pre-trained models to those of two model variants that were trained from scratch \citep[with weights initialized according to the random uniform initialization scheme proposed by][]{glorot_understanding_2010}.
After 50 training epochs, the model variants that were not pre-trained achieved a final decoding accuracy of $80.83\%$ (2D; Fig.\ \ref{fig:results:fine-tuning-performance} A) and $82.23\%$ (3D; Fig.\ \ref{fig:results:fine-tuning-performance} D) in the validation data of the HCP-WM task, thereby performing $-1.27\%$ ($t(49)=-4.047, P=0.00018$) and $-1.22\%$ ($t(49)=-5.74, P<0.0001$) worse than their pre-trained counterparts (Fig.\ \ref{fig:results:fine-tuning-performance} C, F)). 

To also test how the pre-trained and not pre-trained model variants compared when applied to smaller fractions of the full training dataset of the HCP-WM task (N = 400), we trained both variants of each architecture on the data of 1\%, 5\%, 10\%, 20\%, 40\%, 60\%, and 80\% of the full training dataset.
Importantly, we always evaluated the decoding performance of each model on the data of all 50 subjects in the validation dataset.

The pre-trained DeepLight variants consistently achieved higher decoding accuracies than the models that were not pre-trained (Fig.\ \ref{fig:results:fine-tuning-performance} C, F).
Note that the pre-trained models were able to correctly identify the cognitive state underlying $69.34\%$ (2D) and $73.10\%$ (3D) of the fMRI volumes of the validation dataset, when they were trained with only 1\% of the training dataset (equal to a dataset of four subjects).
The DeepLight variants that were not pre-trained, on the other hand, achieved a decoding accuracy of $29.87\%$ (2D) and $25.02\%$ (3D) when trained on 1\% of the training data (thereby performing meaningfully worse than their pre-trained counterparts; the difference in decoding accuracy between the pre-trained and not pre-trained models was $-39.47\%$, ($t(49) = -29.13, P < 0.0001$) and $-48.08\%$ ($t(49)=-30.20, P<0.0001$) for the 2D- and 3D architectures respectively).

Similarly, the pre-trained DeepLight variants that were fine-tuned on 40\% of the training dataset achieved a final decoding performance that was as good as the performance of the DeepLight variants that were trained from scratch on the data of all 400 subjects in the training dataset (2D: $80.83\%$ (not pre-trained; 100\%) $-$  $80.11\%$ (pre-trained; 40\%) = $0.72\%$ ($t(49) = 2.57, P=0.013$), 3D: $82.23\%$ (not pre-trained; 100\%) $-$ $81.96\%$ (pre-trained; 40\%) = $0.27\%$ ($t(49) = 1.58, P=0.12$); Fig.\ \ref{fig:results:fine-tuning-performance} A-B, D-E)

Overall, the 3D-DeepLight variants were slightly more accurate in identifying the cognitive states from the fMRI data than their 2D counterparts (Fig.\ \ref{fig:results:fine-tuning-performance} G - I).
While the 3D-DeepLight variants generally also learned faster (by achieving higher decoding accuracies earlier in the training; Fig.\ \ref{fig:results:fine-tuning-performance} G-H), we refrain from interpreting this finding further, as we used slightly different learning rates to train the two DeepLight architectures (for details on the training procedures, see section \ref{sec:methods:deeplight-training}).

\subsection{Pre-trained models transfer well to an independent fMRI dataset}
\label{sec:deeplight:results:transfer:multi-task-data}

\begin{figure}[!t]
\begin{center}
\includegraphics[width=\linewidth]{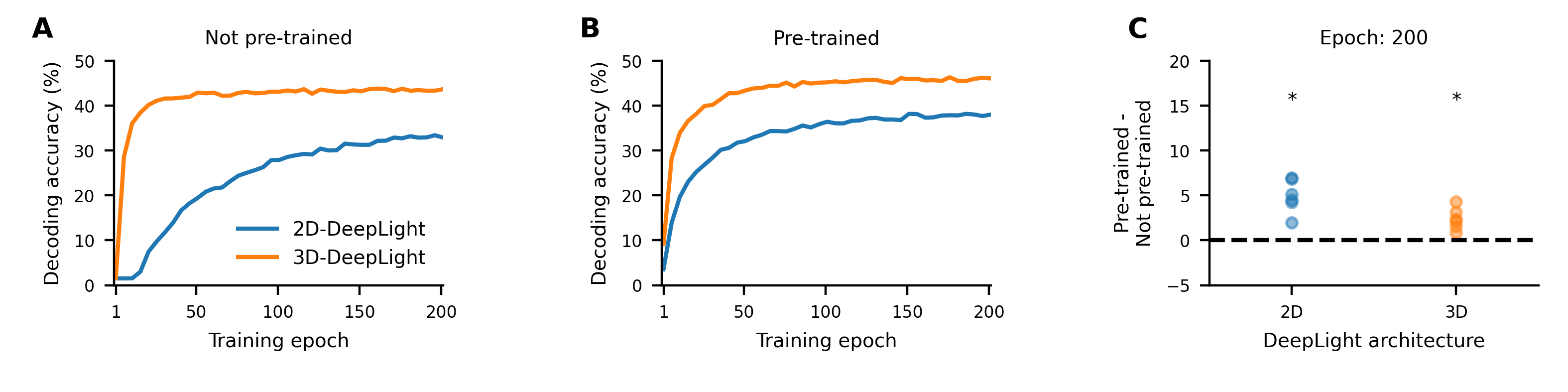}
\end{center}
    \caption{Decoding accuracy of DeepLight variants that are trained from scratch (A) or pre-trained (B) in the validation data of the "Multi-task" dataset (see section \ref{sec:methods:data:multi-task}). A-B: The 3D-DeepLight variants consistently achieve higher decoding accuracies than their 2D counterparts, while the pre-trained model variants also outperform their not pre-trained counterparts.
    For an overview of the training procedures, see section \ref{sec:methods:deeplight-training}.
    An epoch is defined as an entire iteration over the training dataset.
    Lines indicate decoding accuracy, while scatters indicate individual subject decoding accuracies, and colors the 2D- (blue) and 3D-DeepLight (orange) architectures.
    Stars indicate that the distribution of subject decoding accuracies is meaningfully different from 0 in a two-sided t-test (such that $P\leq0.005$).}
\label{fig:deplight:results:transfer:multi-task-dataset}
\end{figure}

Our analyses have shown that the pre-trained DeepLight variants consistently achieve higher decoding accuracies than model variants that were not pre-trained when both are applied to the fMRI data of the HCP-WM task.
To also test whether the pre-trained models exhibit a similar advantage in decoding performance when applied to an independent fMRI dataset that is not part of the HCP, we performed a similar transfer learning analysis on a dataset that was originally published by by Nakai and Nishimoto \citep[the "Multi-task" dataset; ][]{nakai_quantitative_2020}.
In this dataset, six participants repeatedly performed 103 simple naturalistic tasks in the fMRI \citep[e.g., deciding whether the music that is currently being played is Jazz or whether there is a penguin on a presented image; for further details on the tasks and dataset, see section \ref{sec:methods:data:multi-task} and][]{nakai_quantitative_2020}.
In total, the Multi-task dataset contains the fMRI data of 18 runs for each individual and is split into a training dataset (containing the data of 12 runs per individual) and a test dataset (containing the data of the remaining six runs per individual).
In the test runs, participants performed versions of the 103 tasks that were not included in the training runs (for example, by utilizing different music or images).
Similar to our previous analyses, we evaluated the performance of a pre-trained and not pre-trained variant of each DeepLight architecture (2D and 3D; see Fig.\ \ref{fig:deeplight-architectures}) in identifying the 103 tasks (i.e., cognitive states) from the fMRI data of this dataset (for an overview of the training procedures, see section \ref{sec:methods:deeplight-training}). 

Both pre-trained DeepLight variants again outperformed their not pre-trained counterparts at identifying the 103 tasks from the fMRI data (Fig.\ \ref{fig:deplight:results:transfer:multi-task-dataset}), while the 3D-DeepLight variants performed better than the respective 2D-DeepLight variants (Fig.\ \ref{fig:deplight:results:transfer:multi-task-dataset} A-B).
After 200 training epochs, the 2D-DeepLight variant that was not pre-trained achieved a final decoding accuracy of $33.02\%$ in the test runs of the Multi-task dataset (blue line in Fig.\ \ref{fig:deplight:results:transfer:multi-task-dataset} A), while the not pre-trained 3D-DeepLight variant achieved a final decoding accuracy of $43.70\%$ (orange line in Fig.\ \ref{fig:deplight:results:transfer:multi-task-dataset} A).
The pre-trained DeepLight variants, in contrast, achieved a final decoding accuracy of $37.96\%$ (2D; Fig.\ \ref{fig:deplight:results:transfer:multi-task-dataset} B) and $46.10\%$ (3D; Fig.\ \ref{fig:deplight:results:transfer:multi-task-dataset} B) respectively.
The pre-trained variants therefore outperformed their not pre-trained counterparts by $4.95\%$ ($t(5)=6.50, P=0.0013$; 2D) and $2.40\%$ ($t(5)=4.90, P=0.0045$; 3D), while the 3D-DeepLight variants outperformed the 2D variants by $10.68\%$ ($t(5)=9.88, P=0.00018$; not pre-trained) and $8.12\%$ ($t(5)=10.85, P=0.0001$; pre-trained). 

Interestingly, the pre-trained 3D-DeepLight variant did not exhibit the same advantages in learning speed over its not pre-trained counterpart that we observed in our previous analyses, as both exhibited similar increases in decoding accuracy per training epoch (Fig.\ \ref{fig:deplight:results:transfer:multi-task-dataset} A-B).
We thus confirmed in a sequence of additional analyses that the transfer performance of the pre-trained 3D-DeepLight variant to the Multi-task dataset was not affected by any basic differences in the statistical properties, noise or preprocessing between the HCP and Multi-task datasets (see Appendix \ref{sec:appendix:results:noise-tests-multi-task}). 
\subsection{Pre-trained models reuse features when trained on new data}
\label{sec:results:representation-similarities}

\begin{figure}[!t]
\begin{center}
\includegraphics[width=\linewidth]{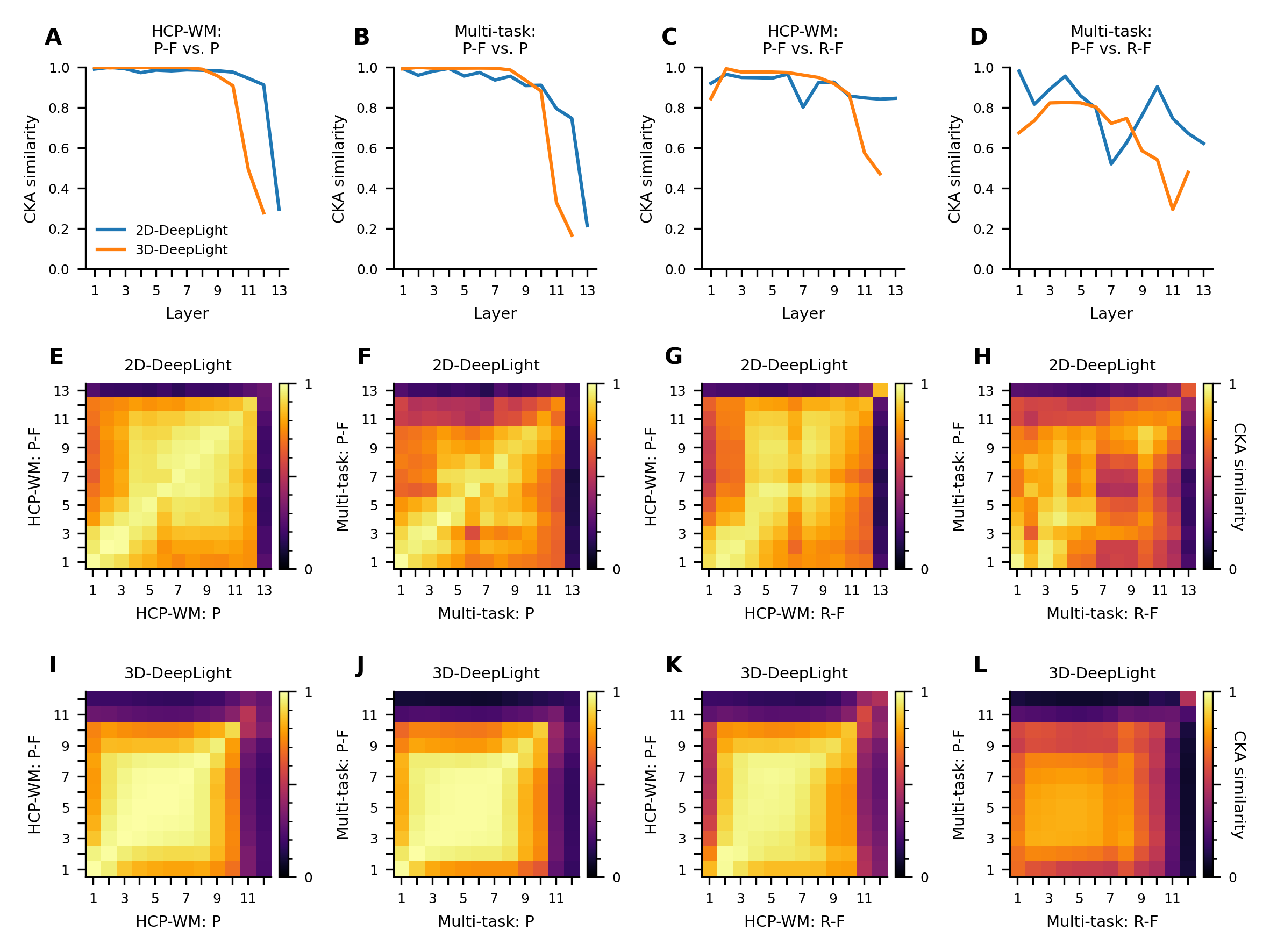}
\end{center}
    \caption{Representational similarity of the hidden layer representations for the validation data of the HCP-WM (A, C, E-I, G-K) and Multi-task datasets (B, D, F-J, H-L) between models that were pre-trained (P), pre-trained and then fine-tuned (P-F), and trained from scratch with randomly initialized weights (R-F).
    Similarity is indicated according to the centered kernel alignment measure \citep[CKA; ][]{kornblith_similarity_2019} using a linear kernel.
    A-B: P and P-F models generally exhibit highly similar hidden layer representations, indicating that the P models reuse many of their learned features during fine-tuning.
    C-D: R-F models, accordingly, exhibit hidden representations that are more dissimilar to those of P-F models.
    E-L: The representation of 2D-DeepLight's LSTM layer (with index 13 in E-H) and the last two layers of 3D-DeepLight's convolutional feature extractor (with index 11 and 12 in I-L) are highly dissimilar between datasets and to the representations of all other model layers.
    Solid lines in A-D and colors in E-L indicate mean CKA values, while the two model architectures in A-D are indicated by blue (2D) and orange (3D).
    For details on the CKA computation, see section \ref{sec:results:representation-similarities}.} 
\label{fig:deplight:results:transfer:representation-similarities}
\end{figure}

To better understand the mechanisms giving rise to the advantages in decoding performance and learning speeds of the pre-trained models that we observed (see Fig.\ \ref{fig:results:fine-tuning-performance} and Fig.\ \ref{fig:deplight:results:transfer:multi-task-dataset}), we next analyzed their hidden layer representations.
We would assume that the pre-trained models are generally able to learn quicker and from less data because they reuse  many of their learned features when training on a new dataset \citep[c.f.,][]{neyshabur_what_2020}.
Accordingly, the hidden layer representations of the pre-trained models should be highly similar for a new dataset before and after training on the dataset.
In contrast, the hidden representations of models that were not pre-trained should be more dissimilar, as their learned features are solely informed by the dataset at hand.

To test these hypotheses, we compared the similarity of the hidden layer representations between three types of models: those that were pre-trained on the data of the six HCP pre-training tasks (see section \ref{sec:results:pre-training}), those that were pre-trained and then fine-tuned on either of the two fine-tuning datasets (namely the HCP-WM and Multi-task datasets; see section \ref{sec:results:fine-tuning}), and those that were trained on these fine-tuning datasets without any pre-training (thus with randomly initialized weights).
In the following, we will indicate these three different model types with P (pre-trained), P-F (pre-trained and fine-tuned), and R-F (trained with randomly initialized weights).

To quantify the similarity between the hidden layer representations of two models, we utilized the centered kernel alignment index (CKA; using a linear kernel), which was recently proposed by \citet{kornblith_similarity_2019} \citep[see also][]{braun_relevant_2008,cristianini_kernel-target_2001,montavon_kernel_2011}.
By the use of this similarity index, we computed a similarity matrix for each model comparison, indicating the similarity between the hidden representations of each pair of layers from the two models (see Fig.\ \ref{fig:deplight:results:transfer:representation-similarities} E-L).
To compute the similarity of two layers' representations, we first extracted their representations of each fMRI volume of a subject in a dataset and then used these representations to compute the CKA similarity for the data of that subject.
To obtain an overall estimate of representation similarity, we next averaged these individual subject similarities over all subjects in a dataset.
Note that we restricted this analysis to only those model layers whose weights were also transferred from the pre-trained models during fine-tuning (thus excluding the output layers of both DeepLight architectures; see Fig.\ \ref{fig:deeplight-architectures} and section \ref{sec:methods:deeplight-architectures}).
We also averaged the representations of convolution layers over their kernel dimension to compensate for their otherwise very high dimensionality. 

By the use of this analysis strategy, we first compared the hidden layer representations of P and P-F models for the data of the HCP-WM (Fig.\ \ref{fig:deplight:results:transfer:representation-similarities} A, E, I) and Multi-task datasets (Fig.\ \ref{fig:deplight:results:transfer:representation-similarities} B, F, J).
This comparison revealed that the hidden representations of P models are generally very similar to those of P-F models in both datasets, indicating that P models reuse many of their learned features during fine-tuning; as we hypothesized.

In contrast, the hidden representations of R-F models are generally more dissimilar to those of P-F models (Fig.\ \ref{fig:deplight:results:transfer:representation-similarities} C-D), demonstrating that R-F models use a more dissimilar set of features to decode the cognitive states of the two datasets; Again, as we hypothesized.

Interestingly, the similarity of the hidden representations in all model comparisons generally decreased for later layers of both DeepLight architectures (i.e., 2D-DeepLight's LSTM layer and 3D-DeepLight's last two layers of the convolutional feature extractor; Fig.\ \ref{fig:deplight:results:transfer:representation-similarities} A-D).
The representations of these layers were also highly dissimilar to the representations of all other model layers (Fig.\ \ref{fig:deplight:results:transfer:representation-similarities} E-L).
This suggests that these higher-level features are more specific to each model and set of cognitive states in a dataset and that they therefore do not generalize as well across datasets as the lower-level features which are closer to the input data.
\subsection{Pre-trained models utilize unforeseen mappings between cognitive states and brain activity}
\label{sec:results:fine-tuning:learned-mapping}

\begin{figure}[!t]
\begin{center}
\includegraphics[width=\linewidth]{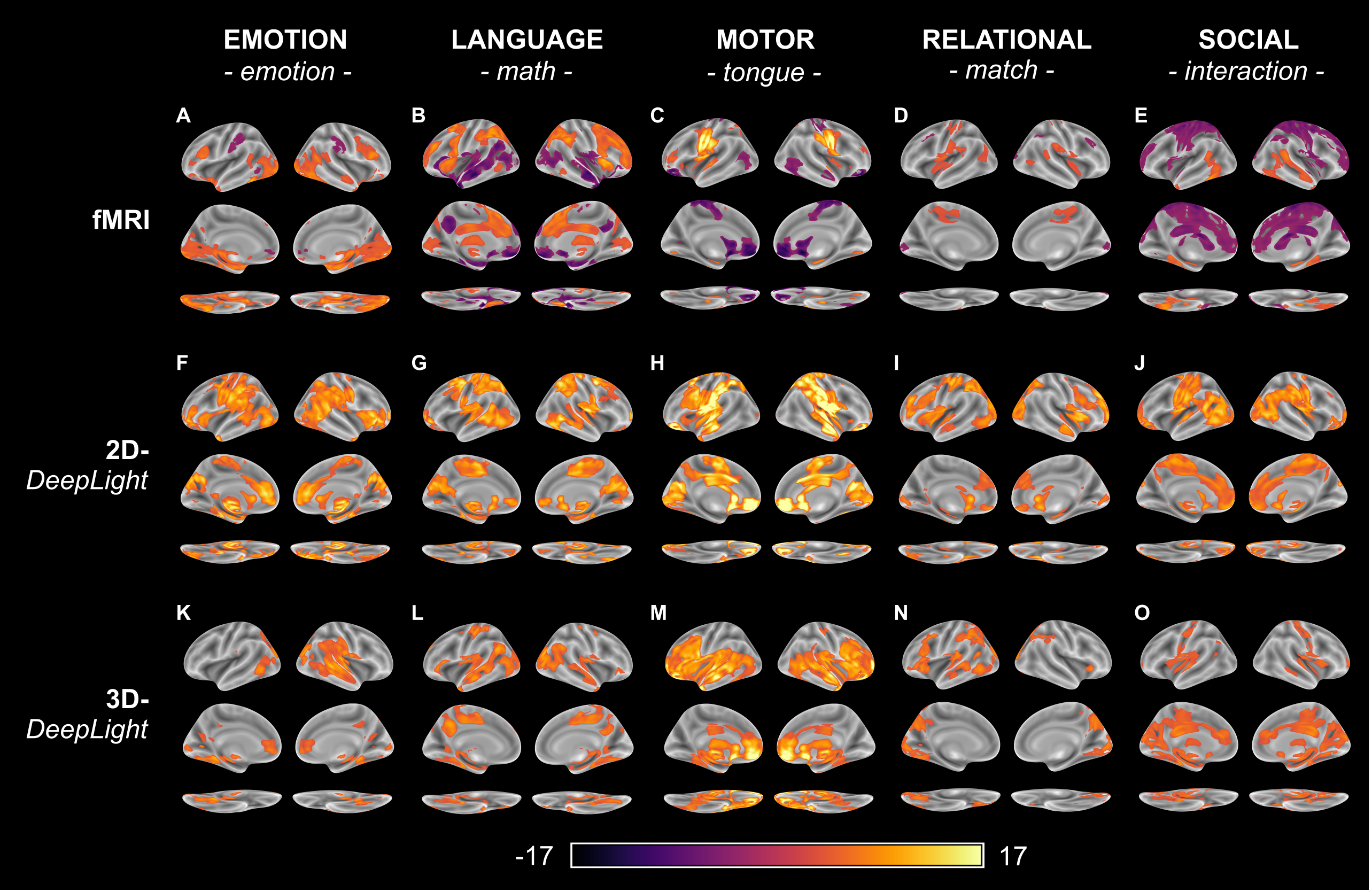}
\end{center}
\caption{Learned mappings between brain activity and cognitive states of the pre-trained DeepLight variants for one exemplary cognitive state of the HCP experimental tasks in the pre-training dataset (excluding the HCP-gambling task; see section \ref{sec:methods:data:hcp}). A-E: For comparison, we first computed a standard two-stage GLM analysis \citep{holmes_generalisability_1998} of the fMRI data of the 50 subjects in the validation dataset of each experimental task. F-O: We also interpreted the decoding decisions of the 2D- (F-J) and 3D-DeepLight (K-O) variants for the same data with the LRP technique (see section \ref{sec:methods:LRP}) and performed a two-stage GLM analysis \citep{holmes_generalisability_1998} of the resulting relevance data to identify the brain regions that each DeepLight variant associated most strongly with a cognitive state (thus restricting the resulting z-scores to positive values). All GLM analyses were performed on parcellated brain data by the use of the dictionaries for functional modes \citep[DiFuMo;][]{dadi_fine-grain_2020} atlas with 256 brain networks and computed separately for each experimental task by contrasting each cognitive state of the task against all other states of that task (for details on the GLM analyses, see Appendix \ref{sec:appendix:methods:glm-analysis-details}). All brain maps are thresholded at a false-discovery rate of 0.001 and projected onto the inflated cortical surface of the FsAverage template \citep{fischl_freesurfer_2012}. Brighter yellow color indicates higher z-scores.}
\label{fig:results:pre-training-brainmaps}
\end{figure}

Our findings demonstrate that the two pre-trained DeepLight variants consistently achieve higher decoding accuracies than their not pre-trained counterparts, while generally also learning quicker and requiring less training data (see sections \ref{sec:results:pre-training}, \ref{sec:results:fine-tuning}, and \ref{sec:deeplight:results:transfer:multi-task-data}). 
Our results also indicate that these advantages arise from the ability of the pre-trained models to reuse many of their learned features when training with new data. 
To better understand the mappings between brain activity and cognitive states that the pre-trained models utilize, we next interpreted their decoding decisions with the LRP technique (for details on the LRP interpretation, see section \ref{sec:methods:LRP}) for the validation data of the pre-training dataset (comprising 50 subjects in each of six HCP experimental tasks; see section \ref{sec:methods:data:hcp}).
For comparison, we also performed a standard two-stage GLM analysis \citep{holmes_generalisability_1998} of the same fMRI data, contrasting each cognitive state of an experimental task against all other states of the task (for details on the GLM analysis, see Appendix \ref{sec:appendix:methods:glm-analysis-details}).
Since LRP interpretation results in a dataset of equal dimension as the original input data (with one relevance value for each input voxel), we performed a similar two-stage GLM analysis of the relevance data of each experimental task to identify the brain regions that the pre-trained models associate most strongly with each cognitive state (thus restricting the results of this analysis to only positive z-scores) (for details on the GLM analysis, see Appendix \ref{sec:appendix:methods:glm-analysis-details}).
All GLM analyses were performed on parcellated brain data, comprising a set of 256 brain networks that were extracted from each fMRI volume by the use of the dictionaries of functional modes (DiFuMo) atlas.
These 256 brain networks are optimized to represent fMRI data well across a wide range of experimental conditions (for methodological details on the DiFuMo atlas, see \citet{dadi_fine-grain_2020}).
Note that we excluded the data of the HCP-gambling task from this analysis, as the pre-trained models did not perform well in decoding the cognitive states of this task (see Fig. \ref{fig:results:pre-training-performance} C,G).

Figure \ref{fig:results:pre-training-brainmaps} provides an overview of the results of this analysis and depicts the brain maps (thresholded at a false-discovery rate of 0.001) for one exemplary cognitive state of each experimental task, namely, the "emotion" state of the HCP-emotion task, in which individuals see images of angry or fearful faces, the "math" state of the HCP-language task, in which individuals solve arithmetic problems, the "tongue" state of the HCP-motor task, in which individuals move their tongue, the "match" state of the HCP-relational task, in which individuals decide if visually presented objects match on a given dimension, and the "interaction" state of the HCP-social task, in which individuals saw objects in a video clip that interacted in some way (for further details on the HCP experimental tasks, see section \ref{sec:methods:data:hcp} and Appendix \ref{sec:appendix:methods:HCP-tasks}). 

In general, the two pre-trained DeepLight variants associated a similar overall set of brain regions with these cognitive states as our GLM analysis of the fMRI data.
Yet, their learned mappings between brain activity and the individual cognitive states are, at first sight, surprising and counterintuitive given the results of the fMRI-GLM analysis (Fig.\ \ref{fig:results:pre-training-brainmaps} A-E).
Take as an example the "tongue" state of the HCP-motor task (Fig.\ \ref{fig:results:pre-training-brainmaps} C,H,M).
The fMRI-GLM analysis (Fig.\ \ref{fig:results:pre-training-brainmaps} C) indicates that the lower parts of the primary motor and somatosensory cortices exhibit overall stronger activation in this state when compared to movements of fingers or toes (the other cognitive states of the HCP-motor ask).
While 2D-DeepLight (Fig.\ \ref{fig:results:pre-training-brainmaps} H) also utilizes the activity of these regions to identify tongue movements, 3D-DeepLight (Fig.\ \ref{fig:results:pre-training-brainmaps} M) does not take the activity of these regions into account when identifying movements of the tongue.
Instead, both DeepLight variants focus more strongly on the activity of the ventromedial prefrontal and anterior cingulate cortices when decoding tongue movements.
These regions are, however, not positively associated with tongue movements in the fMRI-GLM analysis (Fig.\ \ref{fig:results:pre-training-brainmaps} C).
To make sense of this mapping, we need to look back at the results of the fMRI-GLM analysis (Fig.\ \ref{fig:results:pre-training-brainmaps} C), which indicate that the ventromedial prefrontal and anterior cingulate cortices exhibit overall less activation during tongue movements relative to the other movements of the HCP-motor task. 
The two pre-trained DeepLight variants have thus learned to associate the relatively lessened activity of these regions with movements of the tongue.

A similar pattern can be observed for the "math" state of the HCP-language task (Fig.\ \ref{fig:results:pre-training-brainmaps} B,G,L).
The fMRI-GLM analysis (Fig.\ \ref{fig:results:pre-training-brainmaps} B) indicates that the superior temporal gyrus is generally less active when individuals solve arithmetic problems, when compared to answering questions about brief fables (the other cognitive state of the HCP-language task).
Again, both DeepLight variants (Fig.\ \ref{fig:results:pre-training-brainmaps} G,L) interpret this lessened activity as evidence for the presence of the "math" state, as indicated by the generally higher relevance values in this region.
Interestingly, both DeepLight variants do not take into account the generally higher activation of the anterior cingulate cortex in this cognitive state (Fig.\ \ref{fig:results:pre-training-brainmaps} B), demonstrating that the set of brain regions that the models associate with individual cognitive states does not fully overlap with the set of brain regions identified by the fMRI-GLM analysis.

Note that the two DeepLight variants also slightly differ in the specific set of brain regions whose activity they associate with the individual cognitive states. 
For example, when decoding the "interaction" state of the HCP-social task (Fig.\ \ref{fig:results:pre-training-brainmaps} E,J,O),  2D-DeepLight (Fig.\ \ref{fig:results:pre-training-brainmaps} J) focuses on activity of the posterior parts of the superior and medial temporal gyri as well as the lateral occipital cortex, whereas 3D-DeepLight (Fig.\ \ref{fig:results:pre-training-brainmaps} O) does not associate these brain regions with this state and instead focuses on activity of the medial parts of the occipital cortex as well as the posterior cingulate cortex.

Summarizing, this analysis shows that while the two pre-trained DeepLight variants associate a similar overall set of brain regions with the cognitive states of the HCP experimental tasks as a standard GLM analysis of the same data, the specifics of their learned mapping between individual states and brain activity are unforeseen and counterintuitive, as they have learned to identify individual cognitive states through the combined activity of brain regions that are generally less active in these states relative to other cognitive states and regions that are relatively more active \citep[much like a "Clever Hans";][]{lapuschkin_unmasking_2019}.
In addition, our results show that two DL models can differently utilize the activity of the same set of brain regions to decode the same set of cognitive states with similar accuracy.

To ensure that these findings are not specific to the pre-training dataset, we replicated this analysis for the data of the 50 subjects in the validation dataset of the HCP-WM task (see Appendix Fig. \ref{fig:appendix:deeplight:fine-tuning-brainmaps}), using the pre-trained models that were fine-tuned on the full training dataset of that task (see section \ref{sec:results:fine-tuning}).
This analysis revealed similar patterns in the model's learned mappings between brain activity and cognitive states, as both pre-trained DeepLight variants associate a similar overall set of brain regions with the cognitive states of the HCP-WM task as a standard GLM analysis of the fMRI data, while identifying individual cognitive states through the combined activity of brain regions that are relatively more active in these states and regions that are relatively less active.
For example, the fMRI-GLM analysis (Appendix Fig. \ref{fig:appendix:deeplight:fine-tuning-brainmaps} B-C) indicates generally reduced activation of the parahippocampal gyrus (also known as the parahippocampal place area PPA; \citet{heekeren_general_2004}) in the "faces" state, relative to the other cognitive states of the HCP-WM task, and similarly reduced activation of the fusiform gyrus (also known as the fusiform face area FFA; \citet{heekeren_general_2004}) in the "places" state.
In line with this activity pattern, 3D-DeepLight (Appendix Fig. \ref{fig:appendix:deeplight:fine-tuning-brainmaps} J-K) associates the "faces" state with reduced activation of the PPA and the "places" state with reduced activation of the FFA, assigning higher relevance values to these regions when decoding these states.
At first sight, 3D-DeepLight's brain maps therefore seem directly at odds with the results of the fMRI-GLM analysis and a wealth of other empirical findings for this task \citep[e.g.,][]{heekeren_general_2004,haxby_distributed_2000}, which clearly associate stronger activity of the FFA with the viewing of faces and stronger activity of the PPA with the viewing of places.
Only by interpreting 3D-DeepLight's brain maps in the context of the results of the GLM analysis of the underlying fMRI data, we can uncover that this seemingly contradictory mapping is in fact biologically plausible and in line with the activity patterns of the fMRI data.

\section{Discussion}
In this work, we systematically evaluated deep transfer learning for cognitive decoding from whole-brain functional neuroimaging data.
We first trained two distinct DL model architectures on a large whole-brain fMRI dataset of the HCP \citep{ugurbil_pushing_2013} and subsequently evaluated their performance in decoding the cognitive states of an independent experimental task and dataset.
The pre-trained models consistently achieved higher decoding accuracies than model variants that were not pre-trained while generally also learning quicker and requiring less training data, underlining the overall benefits of pre-training. 

Our findings suggest that these advantages result from the ability of the pre-trained models to reuse the features that they have learned during pre-training when training on a new dataset.
This finding is in line with recent empirical work in computer vision, demonstrating that DL models, which start their training from pre-trained weights, generally exhibit highly similar features before and after training on other datasets \citep{neyshabur_what_2020}.
Here, researchers assume that the pre-trained models stay in the same basin of the loss function when trained on new data, leading to similar learned features and overall quicker convergence on a solution to the learning task.
Models that are trained from scratch, on the other hand, start their training with randomly initialized weights and thus generally require more training time to learn a set of viable features, as these features are solely informed by the training dataset at hand.
Our findings generally support these notions, as the pre-trained models achieve higher overall decoding accuracies in the two validation datasets, generally learn quicker, and exhibit highly similar features before and after training on the two validation datasets, when compared to model variants that are not pre-trained.

In our analyses, the feature reuse of the pre-trained models was mostly restricted to lower-level features, which are closer to the input data, as the hidden representations of layers that are deeper in the models differed more strongly between the different applications of the models.
This indicates that these lower-level features generalize better between datasets by providing a more general projection of the fMRI data into a lower-dimensional space, which preserves important variance of the brain activity \citep[c.f.,][]{braun_relevant_2008,montavon_kernel_2011}.
The higher-level features, in contrast, are aimed at  discriminating a specific set of cognitive states from this lower-dimensional representation \citep[e.g., by suppressing features that are not of interest; ][]{haufe_interpretation_2014} and are therefore more specific to a dataset and model.

We also found that the 2D-DeepLight architecture consistently achieved slightly lower decoding accuracies than its 3D counterpart.
2D-DeepLight is inspired by recent work in computer vision for video data \citep[e.g.,][]{donahue_long-term_2017,marban_recurrent_2019}, where the combination of 2D-convolution and recurrent neural network unit is often beneficial over the application of 3D-convolutions to the image sequences.
Here, such 2D-architectures can explicitly account for the contents of each image of a sequence, while also accounting for the changes from each image to the next.
For fMRI data, however, 2D-DeepLight creates an artificial separation between the axial slices of an fMRI volume.
This makes the decoding of cognitive states potentially more challenging than for 3D-DeepLight, which directly operates in the volumetric space of fMRI.

In contrast to other recent empirical work utilizing DL models for the decoding of fMRI data \citep[e.g., ][]{zhang_functional_2021}, DeepLight does not specifically account for the temporal evolution of brain activity, as it processes individual fMRI volumes independently from one another.
In theory, this is a strongly limiting assumption, as understanding the temporal dynamics of brain activity is fundamental to understanding the dynamics of the brain.
To explicitly account for the temporal evolution of brain activity, DeepLight can be extended with another recurrent network layer, which takes as input the higher-level representations of whole-brain activity resulting from any of the two DeepLight architectures used in this work.
Recent empirical work has already demonstrated that 2D-DeepLight's relevance values follow the temporal evolution of the haemodynamic response measured by fMRI \cite{thomas_analyzing_2019}, suggesting that it captures the measured brain signal well at each time point.
The two pre-trained DeepLight architectures that we publish with this work thus provide promising starting points for such an extended DeepLight architecture that specifically accounts for the temporal evolution of the fMRI signal.

Our findings also point towards nuanced challenges for the application of DL models to the decoding of cognitive states from whole-brain functional neuroimaging data.
When interpreting the decoding decisions of the two pre-trained DeepLight variants for the fMRI data of six out of the seven HCP experimental tasks (excluding the HCP-gambling task), we found that both models generally associated a similar overall set of brain regions with the cognitive states of these tasks as a standard two-stage GLM analysis of the fMRI data (see Fig.\ \ref{fig:results:pre-training-brainmaps} and Appendix Fig. \ref{fig:appendix:deeplight:fine-tuning-brainmaps}).
Surprisingly, they learned to identify individual cognitive states through the activity of brain regions that are generally less active in these states in combination with the activity of regions that are generally more active in these states. 
At first sight, the resulting brain maps can therefore seem at odds with the results of the standard GLM analysis of the same data, as the pre-trained models assign high relevance to brain regions that the GLM analysis negatively associates with these states.
Yet, the learned mappings of the pre-trained models are in fact biologically plausible, given the underlying decoding task. 
Cognitive decoding analyses are aimed at characterizing the representations of a brain region by identifying the cognitive states that can be reliably identified from the activity of the region \citep[c.f.][]{norman_beyond_2006}. 
Knowing that a particular brain region is consistently less active in a particular cognitive state (relative to other states) is therefore similarly informative about the representations of this region as knowing that it is consistently more active in other states.

More generally, an interpretation of the cognitive decoding decisions of DL models allows us to identify brain networks whose activity is associated with the decoded cognitive states.
Yet, the interpretation does not necessarily indicate the characteristics of the mapping between the activity of these networks and the cognitive states.
To accurately characterize this mapping, it is essential to compare the results of the interpretation analysis to those of other analysis  approaches for the same data as well as related empirical findings.
DL models are extremely powerful representation learners \citep[for a detailed discussion, see][]{goodfellow_deep_2016,lecun_deep_2015}, which have been shown to often learn mappings between input data and target signal that are not foreseen by the researchers training these models \citep[e.g.,][]{geirhos_shortcut_2020,lapuschkin_unmasking_2019,oakden-rayner_hidden_2020,mccoy_right_2019,zech_variable_2018}. 
Their learned mappings therefore need to be interpreted with caution, when aiming to draw inferences about the association between brain activity and cognitive states.

Relatedly, our analyses have shown that two DL models, which achieve similar overall decoding performances, can differ in how they combine the activity of the same set of brain regions to distinguish the same set of cognitive states.
This suggests that DL models are able to adopt multiple different solutions for the mapping between a set of cognitive states and brain activity that allow them to perform similarly well on average.
While this finding might seem trivial at first, it has far-reaching consequences, as it brings forward questions about the reproducibility of inferences about the mapping between brain activity and cognitive states that are drawn from the interpretation of cognitive decoding decisions of DL models.
Recent empirical work in DL research has demonstrated that the convergence of DL models, and thereby the specifics of their learned mapping between input data and target signal, is dependent on many non-deterministic aspects of the training process, such as random seeds and random weight initializations \citep{henderson_deep_2018,lucic_are_2018,reimers_reporting_2017,dodge_show_2019} as well as the specific choices for other hyper-parameters, such as individual layer specifications and optimization algorithms \cite{melis_state_2017,zoph_neural_2017,lucic_are_2018}.
It is thus possible that the mapping between cognitive states and brain activity that a DL model learns can change with these factors of the training.
In line with other researchers \citep{bouthillier_accounting_2021,dodge_show_2019,thomas_challenges_2021}, we therefore urge researchers who are interested in drawing inferences about the mapping between cognitive states and brain activity from an interpretation of the cognitive decoding decisions of DL models to test for the influence of convergence on the models' learned mappings, for example, by repeatedly training the same model on the same dataset while varying different non-deterministic aspects of the training (e.g., random seeds, random weight initialzations, and random shufflings and augmentations of the training data) as well as other hyper-parameters (e.g., batch sizes, individual layer specifications, learning rates), and to study the effect of these variations on the model's learned mapping between brain activity and cognitive states.

In conclusion, this work clearly underlines the overall benefits of pre-training for the application of DL models to cognitive decoding from whole-brain fMRI data, as pre-trained models generally exhibit higher decoding accuracies and require less training time and data when compared to model variants that are not pre-trained.
Yet, this work also surfaces nuanced challenges that arise for the application of DL models to whole-brain decoding, as the mappings between brain activity and cognitive states that DL models learn can be unforeseen and counterintuitive, and thus require careful additional analysis to be understood well.

\section{Methods}
\subsection{Overview of datasets and experimental tasks}
\label{sec:methods:data}

\subsubsection{Human Connectome Project}
\label{sec:methods:data:hcp}

\begin{table}[!h]
\centering
\begin{tabularx}{\textwidth}{XlXl}
\toprule
Task &  Cognitive states & Count & Duration (min)\\
\midrule
WM & body, face, place, tool & 4 & 5:01\\
Gambling & win, loss, neutral & 3 & 3:12\\
Motor & left/right finger, left/right toe, tongue & 5 & 3:34\\
Language & story, math & 2 & 3:57\\
Social & interaction, no interaction & 2 & 3:27\\
Relational & relational, matching & 2 & 2:56\\
Emotion & emotion, neutral & 2 & 2:16\\
\midrule
\textbf{Total} & & \textbf{20} & \textbf{23:03}\\
\midrule[\heavyrulewidth]
\end{tabularx}
\caption{Overview of the HCP fMRI data. For each experimental task, the cognitive states, number of cognitive states, and duration of the task in each fMRI run are presented.}
\label{table:methods:data:hcp}
\end{table}

The task-fMRI data of the Human Connectome Project \citep{barch_function_2013} includes seven tasks that were each performed in two separate runs (for a general overview, see Table \ref{table:methods:data:hcp}). For each task, participants were first provided with detailed instructions outside of the fMRI and only given a very brief reminder of the task and a refresher on the response button box mappings before the start of each task in the fMRI. We briefly summarize each experimental task in the following, while a more detailed description of the tasks can be found in Appendix \ref{sec:appendix:methods:HCP-tasks}.

\paragraph{Working memory (WM):} Participants were asked to decide in an N-back task whether a currently presented image (of body parts, faces, places or tools) is the same as a previously presented target image. The target image was either presented at the beginning of the experimental block (0-back) or participants were asked to decide whether the currently presented image is the same as the one presented two before (2-back). 

\paragraph{Gambling:} Participants were asked to guess whether the value of a card (with values between 1-9) is below or above 5. Participants won or lost if they guessed correctly/incorrectly. Trials were neutral if the value of the card was 5. The number on the card was dependent on whether the respective trial belonged to the reward, loss, or neutral task condition.

\paragraph{Motor:} Participants were presented with visual cues asking them to tap their left or right fingers, squeeze their left or right toes, or move their tongue.

\paragraph{Language:} Participants either heard a brief fable (story trials) or an arithmetic problem (math trials) and were subsequently given a two-alternative question about the story or arithmetic problem.

\paragraph{Social:} Participants were presented with short video clips of objects that either interacted in some way (interaction trials) or moved randomly  (no interaction trials). Subsequently, participants were asked to decide whether the objects interacted with one another, did not have an interaction, or if they are not sure.

\paragraph{Relational:} Participants were presented with different shapes, filled with different textures. In relational trials, participants saw a pair of objects at the top of the screen and a pair at the bottom. They were then asked to decide whether the bottom pair differs along the same dimension (shape or texture) as the top pair. In match trials, participants saw one object at the top and bottom and were asked to decide whether the objects matched on a given dimension.

\paragraph{Emotion:} In emotion trials, participants were asked to decide which of two faces presented on the bottom of the screen matches the face at the top of the screen (faces had an either angry or fearful expression). In neutral trials, participants were asked to to decide which of two shapes at the bottom of the screen matches a shape that is presented at the top.

\bigbreak

fMRI data for each task were provided in a preprocessed format by the Human Connectome Project (HCP S1200 release), WU Minn Consortium (Principal Investigators: David Van Essen and Kamil Ugurbil; 1U54MH091657) funded by the 16 NIH Institutes and Centers that support the NIH Blueprint for Neuroscience Research; and by the McDonnell Center for Systems Neuroscience at Washington University. Whole-brain EPI acquisitions were acquired with a 32 channel head coil on a modified 3T Siemens Skyra with TR = 720 ms, TE = 33.1 ms, flip angle = 52 deg, BW = 2,290 Hz/Px, in-plane FOV = 208 x 180 mm, 72 slices, 2.0 mm isotropic voxels with a multi-band acceleration factor of 8. Two runs were acquired, one with a right-to-left and the other with a left-to-right phase encoding  \citep[for further methodological details on fMRI data acquisition, see][]{ugurbil_pushing_2013}.

The Human Connectome Project preprocessing pipeline for functional MRI data \citep[“fMRIVolume”;][]{glasser_minimal_2013} includes the following steps: gradient unwarping, motion correction, fieldmap-based EPI distortion correction, brain-boundary based registration of EPI to structural T1-weighted scan, non-linear registration to MNI152 space, and grand-mean intensity normalization \citep[for further details, see][]{glasser_minimal_2013,ugurbil_pushing_2013}.

In addition to the minimal preprocessing of the fMRI data that was performed by the Human Connectome Project, we applied the following preprocessing steps to the fMRI data for our decoding analyses: volume-based smoothing of the fMRI sequences with a 3mm FWHM Gaussian kernel, linear detrending and standardization of the single voxel signal time-series (resulting in a zero-centered voxel time-series with unit variance) and temporal filtering of the single voxel time-series with a butterworth highpass filter and a cutoff of 128 s, as implemented in Nilearn \citep{abraham_machine_2014}.

\subsubsection{Multi-task dataset}
\label{sec:methods:data:multi-task}
This dataset was first published by \citet{nakai_quantitative_2020} and contains the data of six healthy human participants (22-33 yrs, two female, normal vision and hearing, all right-handed) who repeatedly performed 103 simple naturalistic tasks in the fMRI (hence "Multi-task"). These tasks were selected such that they include a variety of different cognitive domains and can be performed without previous experimental training \citep[e.g., participants were asked whether a piece of music is Jazz or whether a penguin is shown on a a presented image; for an overview of all tasks and instructions, see][]{nakai_quantitative_2020}. The experiment consisted of 18 fMRI runs of which 12 were designated as training runs and six as test runs. In each run, 77-83 trials were presented with a duration of 6-12 s per trial. Additionally, a two second feedback (correct, incorrect) for the preceding task was presented 9-13 times per run. Each task had 12 different instances of which eight were used in the training runs and four in the test runs. Importantly, there was no overlap between the training and test instances of each task. The task order was pseudorandomized during the training runs, as some tasks depended on one another. In the six test runs, all tasks were presented in the exact same order. Subjects did not receive any explanation of the tasks prior to the experiment and only underwent a small training on how to use the buttons in the fMRI to indicate their responses (with one response pad with two buttons for each hand). The instruction text of each task was presented with the respective stimuli as a single image during the experiment. All stimuli were shown on a projector screen (21.0 x 15.8° of visual angle at 30 Hz). The experiment was performed over three days, with six runs on each day.

The unprocessed fMRI data for this experiment were obtained from the original authors \citep{nakai_quantitative_2020} through OpenNeuro \citep{markiewicz_openneuro_2021}. Whole-brain EPI acquisitions were acquired with a 32 channel head coil on a 3T Siemens TIM Trio with TR = 2000 ms, TE = 30 ms, flip angle = 62 ° , FOV = 192 x 192 mm, resolution = 2 x 2 mm, MB factor = 3. 72 interleaved axial slices were scanned parallel to the anterior and posterior commissure line (that were each 2.0-mm thick without a gap), using a T2*-weighted gradient-echo multiband echo-planar imaging (MB-EPI) sequence. 275 volumes were obtained for each run. In addition, high-resolution T1-weighted images of the whole brain were acquired from all subjects with a magnetization-prepared rapid acquisition gradient echo sequence (MPRAGE, TR = 2530 ms, TE = 3.26 ms, FA = 9 °, FOV = 256 x 256 mm, voxel size = 1 x 1 x 1 mm).

We preprocessed these data using \emph{fMRIPrep} 20.0.5 (\cite{esteban_fmriprep_2019}; \cite{fmriprep2}; RRID:SCR\_016216), which is based on \emph{Nipype} 1.4.2 (\cite{gorgolewski_nipype_2011}; \cite{nipype2}; RRID:SCR\_002502). All details on the individual preprocessing steps for this dataset are reported in Appendix \ref{sec:appendix:methods:preprocessing-multitask}.

\subsection{DeepLight architectures}
\label{sec:methods:deeplight-architectures}

We used two distinct DL model architectures in this work, which we refer to as 2D- and 3D-DeepLight:

\paragraph{2D-DeepLight}
\label{sec:methods:2D-DeepLight}
The 2D-DeepLight architecture is based on the model used in \citet{thomas_analyzing_2019,thomas_deep_2019} and is composed of three distinct computational modules, namely, a 2D-convolutional feature extractor, an LSTM, and an output layer (for an overview, see Fig.\ \ref{fig:deeplight-architectures}). First, 2D-DeepLight separates each fMRI volume into a sequence of 2D axial brain slices. These slices are then processed by a 2D-convolutional feature extractor \citep{lecun_gradient-based_1998}, resulting in a sequence of higher-level, and lower-dimensional, slice representations. These higher-level slice representations are fed to an LSTM \citep{hochreiter_long_1997}, integrating the spatial dependencies of the observed brain activity within and across the axial slices. Lastly, the output layer makes a decoding decision, by projecting the output of the LSTM into a lower-dimensional space, which spans the cognitive states in the data. Here, a probability for each cognitive state is estimated, indicating whether the fMRI volume belongs to each of these states. This combination of convolutional and recurrent DL elements is inspired by previous research demonstrating that is well-suited to learn the spatial dependency structure of long sequences of input data \citep[e.g.,][]{donahue_long-term_2017,marban_recurrent_2019}.

The 2D-convolutional feature extractor of this 2D-DeepLight variant is composed of a sequence of  2D-convolution layers \citep{lecun_gradient-based_1998}. A 2D-convolution layer consists of a set of kernels (or filters) $w$ that each learns local features of an input image $x$. These local features are then convolved over the input, resulting in an activation map $h$, indicating whether a feature is present at each given location of the input:

\begin{equation}
h_{i,j} = g(\sum_{m=1}^M \sum_{n=1}^N w_{m,n}x_{i+m-1, j+n-1} + b)
\end{equation}

Here, $b$ represents the bias of the kernel, while $g$ represents the activation function (see eq.\ \ref{eq:relu}). The indices $m$ and $n$ represent the row and column index of the kernel matrix, whereas $i$ and $j$ represent the coronal (i.e., row) and saggital (i.e., column) dimensions of the activation map.

Specifically, we used the following sequence of 12  2D-convolution layers for the 2D-convolutional feature extractor (notation: conv(kernel size) - (number of kernels)(stride size)): conv3-8(1), conv3-8(1), conv3-16(2), conv3-16(1), conv3-32(2), conv3-32(1), conv3-32(2), conv3-32(1), conv3-64(2), conv3-64(1), conv3-64(2), conv3-64(1).

Generally, lower-level convolution kernels (closer to the input data) have small receptive fields and are only sensitive to local features of small patches of the input data (e.g., contrasts and orientations). Higher-level convolution kernels, on the other hand, act upon a higher-level representation of the input data, which has already been transformed by a sequence of preceding lower-level convolution kernels. Higher-level kernels thereby integrate the information provided by lower-level convolution kernels, allowing them to identify larger and more complex patterns in the data.

All convolution kernels in 2D-DeepLight are activated through a rectified linear unit function (ReLU):

\begin{equation}
\label{eq:relu}
g(x) = \max(0, x)
\end{equation}

All convolution kernels of the odd-numbered layers are moved over the input fMRI data with a stride size of one voxel while all kernels of even-numbered layers are moved with a stride size of two voxels. An increasing stride indicates more distance between the applications of the convolution kernels to the input data, thereby reducing the dimensionality of the output representation at the cost of a decreasing sensitivity to differences in the activity patterns of neighboring voxels. As the activity patterns of neighboring voxels are known to be highly correlated, the overall loss of information through a reasonable increase in stride size is generally low. We further applied zero-padding to the inputs of each convolution layer, such that the outputs of the convolution layers have the same dimensionality as their inputs, if a stride of one voxel is applied, and only decrease in size when larger strides are used.

To integrate the information provided by the resulting sequence of slice representations into a higher-level representation of the observed whole-brain activity, 2D-DeepLight applies a bi-directional LSTM unit \citep{hochreiter_long_1997}, which contains two independent LSTM units. Each of the two units iterates over the entire sequence of input slices, but in reverse order (one from bottom-to-top and the other from top-to-bottom). An LSTM unit contains a hidden cell state $C$, storing information over an input sequence of length $S$ with elements $x^{(s)}$ and outputs a vector $h^{(s)}$ for each input at sequence step $s$. The unit has the ability to add and remove information from $C$ through a series of gates. In a first step, the LSTM unit decides which information from the cell state $C^{(s-1)}$ is removed. This is done by a fully-connected logistic layer, the forget gate $f$:

\begin{gather}
\label{eq:methods:2d-deeplight:lstm-forget-gate}
f^{(s)} = \sigma(W_f \cdot x^{(s)} + U_f \cdot h^{(s-1)}+b_f) \\
\sigma(x) = \frac{1}{(1+e^{-x})}
\end{gather}

Here, $\sigma$ indicates the logistic function, $[W, U]$ the weight matrices of the forget gate, and $b$ the gate's bias. The forget gate outputs a number between 0 and 1 for each entry in the cell state $C$. Next, the LSTM unit decides which information is stored in the cell state. This operation contains two elements: the input gate $i$, which decides which values of $C^{(s-1)}$ will be updated, and a $tanh$ layer, which creates a new vector of candidate values $C^{\prime(s)}$:

\begin{gather}
\label{eq:methods:2d-deeplight:lstm-input-gate}
i^{(s)} = \sigma(W_i \cdot x^{(s)} + U_i \cdot h^{(s-1)} + b_i) \\
C^{\prime(s)} = tanh(W_c \cdot x^{(s)} + U_c \cdot h^{(s-1)} + b_c) \\
tanh(x) = \frac{e^{x} - e^{-x}}{e^{x} + e^{-x}}
\end{gather}

Subsequently, the old cell state $C^{(s-1)}$ is updated into the new cell state $C^{(s)}$:

\begin{equation}
\label{eq:methods:2d-deeplight:lstm-new-C}
C^{(s)} = f^{(s)} C^{(s-1)} + i^{(s)}C^{\prime(s)}
\end{equation}

Lastly, the LSTM computes its output $h^{(s)}$. Here, the output gate $o$, decides what part of $C^{(s)}$ will be outputted. Subsequently, $C^{(s)}$ is multiplied by another $tanh$ layer to make sure that $h^{(s)}$ is scaled between -1 and 1:

\begin{gather}
\label{eq:methods:2d-deeplight:lstm-output-gate}
o^{(s)} = \sigma(W_o \cdot x^{(s)} + U_o \cdot h^{(s-1)} + b_o) \\
h^{(s)} = o^{(s)} \ tanh(C^{(s)})
\end{gather}

To make a decoding decision, both LSTM units forward their output for the last sequence element $h^{(S)}$ to a fully-connected softmax output layer, with one unit $k$ for each of the $K$ cognitive states in the data:

\begin{equation}
p_k = \frac{e^{h_k^{(S)}}}{\sum_{i=1}^{K}e^{h_i^{(S)}}}
\end{equation}

\paragraph{3D-DeepLight}
\label{sec:methods:3D-DeepLight}
3D-DeepLight replaces the combination of 2D-convolution and LSTM that is used by 2D-DeepLight (see Fig.\ \ref{fig:deeplight-architectures}) with a 3D-convolutional feature extractor that directly accounts for the three spatial dimensions of whole-brain fMRI data.

A 3D-convolution layer consists of a set of 3D-kernels $w$ that each learn specific features of an input volume $x$. In contrast to the features learned by 2D-convolution kernels, these features can be three-dimensional (or volumetric). Similar to 2D-convolutions, these features are convolved over the input, resulting in a set of activation maps $h$, indicating the presence of each of these features at each spatial location of the input volume:

\begin{equation}
h_{i,j,l} = g(\sum_{m=1}^M \sum_{n=1}^N \sum_{z=1}^Z w_{m,n,z}x_{i+m-1, j+n-1, l+z-1} + b)
\end{equation}

Again, $b$ represents the bias of the kernel, while $g$ represents the rectified linear unit activation function (see eq.\ \ref{eq:relu}). The indices $m$, $n$, and $z$ index the row, column, and height of the 3D-convolution kernel, while $i$, $j$, and $l$ indicate the coronal (i.e., row), saggital (i.e., column), and axial (i.e., height) dimension of the activation map $h$.

We used the following sequence of 12 3D-convolution layers for the convolutional feature extractor of 3D-DeepLight (notation: conv(kernel size) - (number of kernels)(stride size)): conv3-8(1), conv3-8(1), conv3-8(2), conv3-8(1), conv3-16(2), conv3-16(1), conv3-32(2), conv3-32(1), conv3-64(2), conv3-64(1), conv3-128(2), conv3-128(1). Similar to 2D-DeepLight, this configuration moves all convolution kernels of the even-numbered layers over the input fMRI volume with a stride size of one voxel and all kernels of odd-numbered layers with a stride size of two voxels, with the exception of the first layer, which applies a stride size of 1 voxel. Similar to 2D-DeepLight, 3D-DeepLight utilizes zero padding, such that the dimensionality of the activation map $h$ only decreases when a stride of more than one voxel is used. 

To make a decoding decision, 3D-DeepLight passes the representation of the feature extractor to an output layer. The output layer is composed of a 1D-convolution layer (with one kernel for each of the cognitive states in the data) as well as a global average pooling layer and softmax function. The purpose of the 1D-convolution layer is to aggregate the information of the $C$ channels of the activation maps $h$, resulting from the 3D-convolutional feature extractor, to one activation map for each of the $K$ cognitive states in the data. These aggregate activation maps are then used to compute one probability estimate $p_k$ for each cognitive state $k$ in the data by the application of global average pooling and softmax scaling:

\begin{gather}
\label{eq:3d-deeplight:output-unit}
h_{i,j,l,k} = g(\sum_{c=1}^{C}w_{k,c}x_{i,j,l,c}) \\
a_k = \frac{1}{MNZ}\sum_{m=1}^M\sum_{n=1}^N\sum_{z=1}^Z h_{m,n,z,k} \\
p_k = \frac{e^{a_k}}{\sum_{i=1}^{K}e^{a_i}}
\end{gather}

Here, $g$ again indicates the rectified linear unit function (see. eq.\ \ref{eq:relu}).

\subsection{DeepLight training}
\label{sec:methods:deeplight-training}
We iteratively trained both DeepLight architectures through backpropagation \citep{hecht-nielsen_theory_1992} by the use of the ADAM optimization algorithm as implemented in tensorflow 1.12 \citep{abadi_tensorflow_2016}. Each training epoch was defined as a complete iteration over all samples in the respective training dataset with a batch size of 32. Weights were initialized by the use of a normal-distributed random initialization scheme \citep{glorot_understanding_2010} (if not noted otherwise).

\paragraph{2D-DeepLight}
 For 2D-DeepLight, we applied dropout regularization to the different model layers \citep{srivastava_dropout_2014} and global gradient norm clipping \citep[with a clipping threshold of 5;][]{pascanu_difficulty_2013} to prevent overfitting. Specifically, we set the dropout probability to 0\% for the first four convolution layers, 20\% for the next four convolution layers, and 40\% for the last four convolution layers during training. We trained 2D-DeepLight with a learning rate of 0.0001.

\paragraph{3D-DeepLight}
Similar to 2D-DeepLight, we applied dropout regularization to 3D-DeepLight's convolution layers \citep{srivastava_dropout_2014} by setting the dropout probability to 20\% during training for these layers. We trained 3D-DeepLight with a learning rate of 0.001.

\subsection{Layer-wise relevance propagation}
\label{sec:methods:LRP}

To relate the decoded cognitive state and brain activity, we here used the layer-wise relevance propagation technique \citep[LRP;][]{bach_pixel-wise_2015,montavon_explaining_2017}. The goal of LRP is to identify the contribution of each  dimension $d$ of an input $x \in {\rm I\!R}^D$  to prediction $f(x)$ of a linear or non-linear predictive function $f$. In the following, the contribution of a single input dimension $d$ to the decoding decision is denoted by its relevance $R_d$. The prediction $f(x)$ can then be decomposed into the sum of the relevance values of each dimension $R_d$ of the input:

\begin{equation}
\label{eq:lrp:sum}
f(x) = \sum_{d=1}^DR_d
\end{equation}

Importantly, LRP assumes that $f(x)>0$ indicates evidence for the presence of a target, while  $f(x)<0$ indicates evidence against the presence of the target. Accordingly, any $R_d>0$ can qualitatively be interpreted as evidence in the data that $f$ sees in favor of the presence of the target, while $R_d<0$ indicates evidence that $f$ sees against the presence of the target. 

The relevance of the output unit at the last model layer $L$ is indicated by $R_d^{(L)}$ and the dimension-wise relevance scores on the input units are indicated by $R_d^{(1)}$. The relevance of any other unit $j$ in any other layer $l$ can then be decomposed into the relevance contributions $R_{i \leftarrow j}^{(l-1, l)}$ of those units $i$ in layer $(l-1)$ which provide inputs to unit $j$ in layer $(l)$:

\begin{equation}
R_j^{(l)} = \sum_{i \in (l-1)} R_{i \leftarrow j}^{(l-1, l)}
\end{equation}

To satisfy eq.\ \ref{eq:lrp:sum}, any definition of the relevance contributions $R_{i \leftarrow j}^{(l-1, l)}$ needs to satisfy the following relevance conservation property between layers $(l)$ and $(l-1)$:

\begin{gather}
\label{eq:lrp:conservation}
\sum_{j \in (l)}R_j^{(l)} = \sum_{i \in (l-1)} R_{i}^{(l-1)}
\end{gather}

For an overview of the different rules that have been proposed to define the relevance contributions $R_{i \leftarrow j}^{(l-1, l)}$, see \citet{bach_pixel-wise_2015}, \citet{kohlbrenner_towards_2020}, and \citet{samek_explaining_2021}. 

Note that in a linear network $f(x)= \sum_i x_{ij}$, in which $R_j = f(x)$, the relevance contributions $R_{i \leftarrow j}$ are directly given by $R_{ij} = x_{ij}$. However, in most DL models, the unit activation follows a non-linear function $\sigma$ of $x_j$, such that $f(x) = \sigma(\sum_i x_{ij})$. Importantly, for two of the prominent activation functions, namely, the rectified linear unit function (see eq. \ref{eq:relu}) and hyperbolic tangent, the pre-activations $x_{ij}$ provide a sensible way of measuring the contribution of each unit $x_i$ to $R_j$ \citep[for a more detailed discussion on this issue, see][]{bach_pixel-wise_2015}.

\paragraph{LRP for 2D-DeepLight}
In the context of this work, and in line with the recommendations by \citet{arras_explaining_2017,arras_what_2017}, the contributions $R_{i \leftarrow j}^{(l-1, l)}$ for all weighted connections of 2D-DeepLight (see, for example, eq.\ \ref{eq:methods:2d-deeplight:lstm-forget-gate}, \ref{eq:methods:2d-deeplight:lstm-input-gate}, \ref{eq:methods:2d-deeplight:lstm-output-gate}) are defined as:

\begin{equation}
\label{eq:lrp:epsilon}
R_{i \leftarrow j}^{(l-1, l)} = \frac{z_{ij}^{(l-1)}}{z_j^{(l-1)} + \epsilon \ sign(z_j^{(l-1)}) } R_{j}^{(l)}
\end{equation}

Here, $z_{ij}^{(l-1)} = x_i^{(l-1)}w_{ij}^{(l-1)}$ (with $w$ indicating the weights and $x$ the input of layer $(l-1)$) and $z_j^{(l-1)} = \sum_iz_{ij}^{(l-1)} + b_j^{(l-1)}$ (with $b_j^{(l-1)}$ indicating the bias of unit $j$). $\epsilon$ represents a stabilizer term that is necessary to avoid numerical degenerations when $z_j^{(l-1)}$ is close to 0, which we set $\epsilon = 0.001$ \citep[in line with][]{thomas_analyzing_2019}. 

Importantly, the LSTM unit of 2D-DeepLight also applies another, multiplicative type of connection (see eq. \ref{eq:methods:2d-deeplight:lstm-new-C} and \ref{eq:methods:2d-deeplight:lstm-output-gate}). Let $z_j^{(l)}$ be an upper-layer unit whose value in the forward pass is computed by multiplying two lower-layer unit values $z_g^{(l-1)}$ and $z_s^{(l-1)}$ such that $z_j^{(l)} = z_g^{(l-1)} z_s^{(l-1)}$. These multiplicative connections occur when one multiplies the outputs of a gate unit, whose values range between 0 and 1, with an instance of the hidden cell state, which we refer to as source unit in the following. For these types of connections, we set the relevances of the gate unit $R_g^{(l-1)} = 0$ and the relevances of the source unit $R_s^{(l-1)} = R_j^{(l)}$, where $R_j^{(l)}$ denotes the relevances of the upper layer unit $z_j^{(l)}$ \citep[as proposed in][]{arras_explaining_2017}. The reasoning behind this rule is that the gate unit already decides in the forward pass how much of the information contained in the source unit should be retained to make the classification; $z_g^{(l-1)}$ thereby controls how much relevance will be attributed to  $z_j^{(l)}$ from upper-layer units. Therefore, while this redistribution scheme seems to ignore the values of the units $z_g^{(l-1)}$ and $z_s^{(l-1)}$ for the redistribution of relevance, these are actually taken into account when computing the value $R_j^{(l)}$ from the relevances of the units of the next upper layers to which $z_j^{(l)}$ is connected.

\paragraph{LRP for 3D-DeepLight}
3D-DeepLight represents a fully-convolutional neural network, in which the convolution kernels are activated through ReLU functions (see eq. \ref{eq:relu}). Based on recent empirical work in computer vision \citep{kohlbrenner_towards_2020}, which has shown that class discriminability and object localization of the LRP technique can be increased for these types of networks, we define the relevance contributions $R_{i \leftarrow j}^{(l-1, l)}$ of all weighted connections of unit $i$ in layer $(l-1)$ to unit $j$ in layer $(l)$ as follows:

\begin{equation}
\label{eq:lrp:alpha-beta}
R_{i \leftarrow j}^{(l-1, l)} = (\alpha\frac{z_{ij}^{(l-1)(+)}}{z_j^{(l-1)(+)}} + \beta\frac{z_{ij}^{(l-1)(-)}}{z_j^{(l-1)(-)}} ) R_j^{(l)}
\end{equation}

Here, $z_j^{(l-1)(+)} = \sum_iz_{ij}^{(l-1)(+)}+b_j^{(l-1)(+)}$ and $z_j^{(l-1)(-)} = \sum_iz_{ij}^{(l-1)(-)}+b_j^{(l-1)(-)}$, where "$+$" and "$-$" indicate the respective positive and negative parts of $z_{ij}^{(l-1)}$ and $b_j^{(l-1)}$. $\alpha$ and $\beta$ represent two weighting parameters, which allow to scale the contribution of $z_{j}^{(l-1)(+)}$ and $z_{j}^{(l-1)(-)}$ to $R_{i \leftarrow j}^{(l-1, l)}$. To satisfy the local conservation property (see eq.\ \ref{eq:lrp:conservation}) $\alpha$ and $\beta$ are restricted to $\alpha + \beta = 1$ (we set $\alpha = 2$, in line with \citet{kohlbrenner_towards_2020,samek_explaining_2021}).

Note that the average pooling layer of the output unit (see eq.\ \ref{eq:3d-deeplight:output-unit}) is a special case of a linearly weighted connection and is thereby subject to the relevance attribution rule defined in eq.\ \ref{eq:lrp:epsilon}.

\subsection{Code availability}
The code and parameter estimates of the pre-trained DeepLight architectures can be found at: https://github.com/athms/evaluating-deeplight-transfer

All data analyses were performed in Python 3.6.8 (Python Software Foundation) by the use of the SciPy 1.5.4 \citep{virtanen_scipy_2020}, NumPy 1.19.5 \citep{oliphant_guide_2015}, Matplotlib 3.3.4 \citep{hunter_matplotlib_2007}, Pandas 1.1.5 \citep{mckinney_data_2010}, Nilearn 0.8.0 \citep{abraham_machine_2014}, Tensorflow 1.12 \citep{abadi_tensorflow_2016}, interprettensor (https://github.com/VigneshSrinivasan10/interprettensor), and iNNvestigate 1.0.8 \citep{alber_innvestigate_2019} packages.
\subsection{Data availability}
The data that support the findings of this study are openly available at the ConnectomeDB S1200 Project page of the HCP (https://db.humanconnectome.org/data/projects/HCP1200) as well as through OpenNeuro (https://openneuro.org/datasets/ds002306). No experimental activity involving the human participants took place at the authors’ institutions.  Only de-identified, publicly released data were used in this study.
\subsection{Ethics statement}

\paragraph{HCP:} Scanning protocols involving human participants were reviewed and approved by Washington University in St. Louis’s Human Research Protection Office (HRPO), IRB 201204036. The participants provided their written informed consent to participate in the study. 

\paragraph{Multi-task:} The experiment was approved by the ethics and safety committee of the National Institute of Information and Communications Technology in Osaka, Japan. All participants provided written consent prior to their participation in the study. 

\bibliographystyle{apalike}
\bibliography{references,fmriprep_references}

\medskip

{\bf Acknowledgements: }
This work was funded by the German Ministry for Education and Research as BIFOLD -- Berlin Institute for the Foundations of Learning and Data (ref.\ 01IS18025A and ref.\ 01IS18037A), the German Research Foundation (DFG) as Math+: Berlin Mathematics Research Center  (EXC 2046/1, project-ID: 390685689), and the European Union’s Horizon 2020 programme (grant no.\ 965221). This work was partly supported by the Institute of Information \& Communications Technology Planning \& Evaluation (IITP) grants funded by the Korea Government (No. 2019-0-00079, Artificial Intelligence Graduate School Program, Korea University and No. 2017-0-00451, Development of BCI based Brain and Cognitive Computing Technology for Recognizing User’s Intentions using Deep Learning). Armin W. Thomas is supported by Stanford Data Science. Correspondence for this article should be addressed to A.W.T., K.R.M., and W.S.
\medskip

{\bf Author contributions: }
A.W.T., K.R.M., and W.S. conceived of DeepLight. A.W.T. implemented all visualizations of the experimental procedures, performed all formal data analyses, wrote the required software, and wrote the original draft of the manuscript. A.W.T., K.R.M., U.L., and W.S. reviewed and edited the manuscript. Funding for this work was acquired by K.R.M. The work was supervised by K.R.M. and W.S.
\medskip

{\bf Competing interests: }
The authors declare no competing interests.

\clearpage
\newpage
\counterwithin{figure}{section}
\counterwithin{table}{section}
\appendix
\section{Supplement to methods}
\subsection{HCP experimental task details}
\label{sec:appendix:methods:HCP-tasks}

\paragraph{Working Memory}
Each of the two runs of the working memory task consisted of eight task (25 s each) and four fixation blocks (15 s each). In each of the task blocks, particpants saw images of one of four different stimulus types (namely, images of body parts, faces, places or tools). These four stimulus types are known to reliably engage distinct cortical regions \citep{downing_cortical_2001} across subjects \cite{peelen_within-subject_2005} and time \citep{fox_defining_2009}. Half of the task blocks used a 2-back working memory task (participants were asked to respond “target” when the current stimulus was the same as the stimulus 2 back) and the other half a 0-back working memory task (a target stimulus was presented at the beginning of each block and participants were asked to respond “target” whenever the target stimulus was presented in the block). Each task block consisted of 10 trials (2.5 s each). In each trial, a stimulus was presented for 2 s followed by a 500 ms interstimulus interval (ISI). We were not interested in identifying any effect of the N-back task condition on the evoked brain activity and therefore pooled the data of both N-back conditions.

\paragraph{Gambling}
Participants played a card guessing game in which they were asked to guess the number on a mistery card. The potential card numbers ranged from 1 to 9 and participants were asked to indicate whether they think that the number is going to be above or below 5. Participants received feedback in form of a number on the card. Importantly, the number on the card was dependent on whether the respective trial was marked as a reward, loss, or neutral trial. In addition to the number, the feedback included a green arrow pointing upwards with "$1$" for reward trials or a red arrow pointing downwards next to "$-0.50$" for loss trials or the number "5" and a gray double headed arrow for neutral trials. Participants had 1.5 s to indicate a guess (during this time a "?" was presented), while the subsequent feedback was presented for 1.0 s. In addition, there was a 1.0 s intertrial interval with a "+" on the screen. The task was presented in blocks that each included eight trials that were either mostly reward (6 rewards trials that were pseudo-randomly interleaved with either 1 neutral and 1 loss trial, 2 neutral trials, or 2 loss trials) or mostly loss (6 loss trials interleaved with either 1 neutral and 1 reward trial, 2 neutral trials, or 2 reward trials) trials. In each of the two fMRI runs there were  2 mostly reward and 2 mostly loss blocks, interleaved with 4 fixation blocks (15 s each, during which a "+" is presented on the screen). All participants were provided with money as a result of completing the experiment. The amount they received was standardized due to the fixed nature of the experiment. 

\paragraph{Motor}
Participants were presented with visual cues that asked them to tap their left or right fingers, squeeze their left or right toes, or move their tongue. The task was presented in blocks of 12 s that each included only one movement type (10 movements). Each block was preceded by a 3 s cue. In each of the two fMRI runs, 13 blocks were presented with 2 blocks for tongue movements, 4 blocks for hand movements (2 left, 2 right), and 4 blocks for foot movements (again, 2 left and 2 right). In addition, three 15 s fixation blocks were included in each run.

\paragraph{Language}
This task consisted of two runs that each interleaved 4 blocks of a story task and 4 blocks of a math task. In the story task, participants were presented with brief auditory stories (5-9 sentences) that were adapted from Aesop's fables. After each story, a 2-alternative forced-choice question asked the particpant about the topic of the story. In the math task, particpants were similarly presented with an auditory math problem that asked them to complete 2-alternative forced choice addition or subtraction problems. For example, participants heard the operation "fourteen plus twelve", followed by "equals" and then two choice alternatives ("twenty-nine or twenty-six"). Participants indicated with a button press whether they choose the first or second answer. The lengths of the blocks varied (with an average of approximately 30 s per block), but the task was designed in such a way that the math task blocks matched the length of the story task blocks (with some additional math trials at the end of a block if needed to complete the 3.8min run).

\paragraph{Social}
Participants were presented with video clips (20 s each) that showed objects (squares, circles, triangles) that either interacted in some way or were moving randomly. After each video clip, participants indicated whether they think that the objects had a social interaction (an interaction that appears as if the objects are taking into account each other's feelings or thoughts), they are not sure, or they think the objects did not interact. Each of the two fMRI runs included 5 video blocks (2 with interaction and 3 without in one run and 3 with interaction and 2 without in the other run) as well as 5 15 s fixation blocks.

\paragraph{Relational}
In this task, participants saw stimuli that were composed of six different shapes that were filled with one of six different textures. In the relational task condition, 2 pairs of objects were presented, one at the top of the screen and the other at the bottom. Participants were told that they should first decide what dimension (shape or texture) differs across the top pair of objects and then whether the bottom pair of objects differs along the same dimension. In the matching condition, participants were shown two objects at the top of the screen and one at the bottom. A word in the middle of the screen then indicated whether participants should decide if the bottom object matched either of the two top objects on the "shape" or "texture" dimension. In the relational condition, stimuli were presented for 3500 ms, with a 500 ms intertrial interval and four trials per block. In the matching condition, stimuli were presented for 2800 ms, with a 400 ms intertrial interval, and a total of five trials per block. Each block lasted a total of 18 s. In each of the two fMRI runs three relational blocks, three matching blocks and three fixation blocks (16 s each) were presented.

\paragraph{Emotion}
In emotion trials, participants were presented with with two faces at the bottom of the screen and one face at the top. These faces had an either angry or fearful expression. The participants were asked to decide which of the two faces on the bottom matches the face at the top. In neutral trials, participants were asked to decide which of two shapes at the bottom of the screen matches a shape that is presented at the top. In this task, trials were presented in blocks of six trials of the same task (face or shape). In each trial, the stimulus was presented for 2 s in addition to a 1 s intertrial interval. Each block was further preceded by a 3 s cue for the task (shape or face). Each of the two fMRI runs included three face and three shape blocks. Due to a bug in the experiment script, the experiment stopped before the final three trials of the last block of each trial (for further details on this bug, see \cite{barch_function_2013}).

\subsection{GLM analysis details}
\label{sec:appendix:methods:glm-analysis-details}

\paragraph{FMRI}
Our GLM subject-level analyses of the fMRI data included one predictor for each of the four cognitive states in the design matrix (each representing a box-car function for the occurrence of a cognitive state). We convolved these predictors with a canonical glover haemodynamic response function \citep[HRF;][]{lindquist_modeling_2009} as implemented in Nilearn 0.8.0 \citep{abraham_machine_2014}, to generate the model predictors. We added temporal derivative terms derived from each predictor, an intercept and an indicator of the experiment run to the design matrix, which we all treated as confounds of no interest. The derivative terms were computed by the use of the cosine drift model as implemented in Nilearn 0.8.0 \citep{abraham_machine_2014}.

To generate a set of group-level brain maps with the GLM, we computed a second-level GLM contrast by the use of the standard two-stage procedure for a random-effects group-level analysis, as proposed by \citet{holmes_generalisability_1998}. Here, the subject-level regression coefficients $\beta$ are treated as random effects in a second-level linear contrast analysis, where the distribution of first-level $\beta$-contrasts is assessed. Contrasts were computed between each cognitive state and all others. The resulting group-level brain maps show the z-scores resulting from this test.

\paragraph{Relevances}
Our GLM analyses of the relevance data resulting from the application of the LRP technique to DeepLight's decoding decisions (for an overview of the LRP technique, see section \ref{sec:methods:LRP} of the main text) included one predictor for each of the four cognitive states in the data (each representing a box-car function for the occurrence of a cognitive state). Our previous analyses have indicated that DeepLight's relevance data show a similar temporal evolution as the HRF \citep[see Fig.\ 6 of ][]{thomas_analyzing_2019}. For this reason, we next convolved the predictors with a canonical glover HRF \citep{lindquist_modeling_2009}, as implemented in Nilearn 0.8.0 \citep{abraham_machine_2014}, to generate a set of model predictors. 

We further added temporal derivative terms derived from each predictor, an intercept and an indicator of the experiment run to the design matrix. The temporal derivative terms were computed by the use of the cosine drift model as implemented in Nilearn 0.8.0 \citep{abraham_machine_2014}. Additionally, we added one regressor to the design matrix indicating the total sum of relevance values contained in each fMRI volume (i.e., TR), to account for the variability in the sum of relevance values between TRs resulting from variability in the certainty of DeepLight's predictions (for an overview of the LRP technique, see section \ref{sec:methods:LRP} of the main text). To also account for non-linear relationships between this regressor and the relevance values, we added regressors for the first derivative of the relevance sums, the squared relevance sums, and the first derivative of the squared relevance sums to the design matrix. All of these predictors were treated as confounds of no interest.

Lastly, we added two regressors to the design matrix indicating whether DeepLight correctly or incorrectly identified the cognitive state of each TR (again in form of two box-car functions). Importantly, we included these two predictors in each computed contrast, by contrasting each cognitive state against all other states and by contrasting correct versus incorrect predictions (e.g., to compute a contrast for the body state in the HCP-WM task (see section \ref{sec:methods:data:hcp} of the main text), we would set the contrast vector to: 3, -1, -1, -1, 1, -1 for the predictors: body, face, place, tool, correct, incorrect). 

To generate a set of group-level brain maps with the GLM, we computed a second-level GLM contrast by the use of the standard two-stage procedure for a random-effects group-level analysis, as proposed by 
\citet{holmes_generalisability_1998}. Here, the subject-level regression coefficients $\beta$ are treated as random effects in a second-level linear contrast analysis, where the distribution of first-level $\beta$-contrasts is assessed. The resulting group-level brain maps show the Z-values resulting from this test.

\subsection{FMRIPrep details for Multi-task data}
\label{sec:appendix:methods:preprocessing-multitask}

This dataset was processed using \emph{fMRIPrep} 20.0.5 (\citet{esteban_fmriprep_2019}; \citet{fmriprep2}; RRID:SCR\_016216), which is based on \emph{Nipype} 1.4.2 (\citet{gorgolewski_nipype_2011}; \citet{nipype2}; RRID:SCR\_002502). 

\begin{description}
\item[Anatomical data preprocessing]
The T1-weighted (T1w) image was corrected for intensity non-uniformity
(INU) with \texttt{N4BiasFieldCorrection} \citep{n4}, distributed with
ANTs 2.2.0 \citep[RRID:SCR\_004757]{ants}, and used as T1w-reference
throughout the workflow. The T1w-reference was then skull-stripped with
a \emph{Nipype} implementation of the \texttt{antsBrainExtraction.sh}
workflow (from ANTs), using OASIS30ANTs as target template. Brain tissue
segmentation of cerebrospinal fluid (CSF), white-matter (WM) and
gray-matter (GM) was performed on the brain-extracted T1w using
\texttt{fast} \citep[FSL 5.0.9, RRID:SCR\_002823,][]{fsl_fast}.
Volume-based spatial normalization to two standard spaces
(MNI152NLin6Asym, MNI152NLin2009cAsym) was performed through nonlinear
registration with \texttt{antsRegistration} (ANTs 2.2.0), using
brain-extracted versions of both T1w reference and the T1w template. The
following templates were selected for spatial normalization: \emph{FSL's
MNI ICBM 152 non-linear 6th Generation Asymmetric Average Brain
Stereotaxic Registration Model} {[}\citet{mni152nlin6asym},
RRID:SCR\_002823; TemplateFlow ID: MNI152NLin6Asym{]}, \emph{ICBM 152
Nonlinear Asymmetrical template version 2009c}
{[}\citet{mni152nlin2009casym}, RRID:SCR\_008796; TemplateFlow ID:
MNI152NLin2009cAsym{]},

\item[Functional data preprocessing]
For each of the 18 BOLD runs found per subject (across all tasks and
sessions), the following preprocessing was performed. First, a reference
volume and its skull-stripped version were generated using a custom
methodology of \emph{fMRIPrep}. Susceptibility distortion correction
(SDC) was omitted. The BOLD reference was then co-registered to the T1w
reference using \texttt{flirt} \citep[FSL 5.0.9,][]{flirt} with the
boundary-based registration \citep{bbr} cost-function. Co-registration
was configured with nine degrees of freedom to account for distortions
remaining in the BOLD reference. Head-motion parameters with respect to
the BOLD reference (transformation matrices, and six corresponding
rotation and translation parameters) are estimated before any
spatiotemporal filtering using \texttt{mcflirt} \citep[FSL
5.0.9,][]{mcflirt}. BOLD runs were slice-time corrected using
\texttt{3dTshift} from AFNI 20160207 \citep[RRID:SCR\_005927]{afni}. The
BOLD time-series (including slice-timing correction when applied) were
resampled onto their original, native space by applying the transforms
to correct for head-motion. These resampled BOLD time-series will be
referred to as \emph{preprocessed BOLD in original space}, or just
\emph{preprocessed BOLD}. The BOLD time-series were resampled into
standard space, generating a \emph{preprocessed BOLD run in
MNI152NLin6Asym space}. First, a reference volume and its skull-stripped
version were generated using a custom methodology of \emph{fMRIPrep}.
Several confounding time-series were calculated based on the
\emph{preprocessed BOLD}: framewise displacement (FD), DVARS and three
region-wise global signals. FD and DVARS are calculated for each
functional run, both using their implementations in \emph{Nipype}
\citep[following the definitions by][]{power_fd_dvars}. The three global
signals are extracted within the CSF, the WM, and the whole-brain masks.
Additionally, a set of physiological regressors were extracted to allow
for component-based noise correction \citep[\emph{CompCor},][]{compcor}.
Principal components are estimated after high-pass filtering the
\emph{preprocessed BOLD} time-series (using a discrete cosine filter
with 128s cut-off) for the two \emph{CompCor} variants: temporal
(tCompCor) and anatomical (aCompCor). tCompCor components are then
calculated from the top 5\% variable voxels within a mask covering the
subcortical regions. This subcortical mask is obtained by heavily
eroding the brain mask, which ensures it does not include cortical GM
regions. For aCompCor, components are calculated within the intersection
of the aforementioned mask and the union of CSF and WM masks calculated
in T1w space, after their projection to the native space of each
functional run (using the inverse BOLD-to-T1w transformation).
Components are also calculated separately within the WM and CSF masks.
For each CompCor decomposition, the \emph{k} components with the largest
singular values are retained, such that the retained components' time
series are sufficient to explain 50 percent of variance across the
nuisance mask (CSF, WM, combined, or temporal). The remaining components
are dropped from consideration. The head-motion estimates calculated in
the correction step were also placed within the corresponding confounds
file. The confound time series derived from head motion estimates and
global signals were expanded with the inclusion of temporal derivatives
and quadratic terms for each \citep{confounds_satterthwaite_2013}.
Frames that exceeded a threshold of 0.5 mm FD or 1.5 standardised DVARS
were annotated as motion outliers. All resamplings can be performed with
\emph{a single interpolation step} by composing all the pertinent
transformations (i.e.~head-motion transform matrices, susceptibility
distortion correction when available, and co-registrations to anatomical
and output spaces). Gridded (volumetric) resamplings were performed
using \texttt{antsApplyTransforms} (ANTs), configured with Lanczos
interpolation to minimize the smoothing effects of other kernels
\citep{lanczos}. Non-gridded (surface) resamplings were performed using
\texttt{mri\_vol2surf} (FreeSurfer).
\end{description}

Many internal operations of \emph{fMRIPrep} use \emph{Nilearn} 0.6.2
\citep{abraham_machine_2014}, mostly within the functional
processing workflow. For more details of the pipeline, see
\href{https://fmriprep.readthedocs.io/en/latest/workflows.html}{the
section corresponding to workflows in \emph{fMRIPrep}'s documentation}.

The above boilerplate text was automatically generated by fMRIPrep with
the express intention that users should copy and paste this text into
their manuscripts \emph{unchanged}. It is released under the
\href{https://creativecommons.org/publicdomain/zero/1.0/}{CC0} license.
\subsection{FMRIPrep details for HCP working memory task}
\label{sec:appendix:preprocessing-fmrirpep-HCP-WM}

Results included in this manuscript come from preprocessing performed
using \emph{fMRIPrep} 20.0.5 (\citet{esteban_fmriprep_2019}; \citet{fmriprep2};
RRID:SCR\_016216), which is based on \emph{Nipype} 1.4.2
(\citet{gorgolewski_nipype_2011}; \citet{nipype2}; RRID:SCR\_002502).

\begin{description}
\item[Anatomical data preprocessing]
The T1-weighted (T1w) image was corrected for intensity non-uniformity
(INU) with \texttt{N4BiasFieldCorrection} \citep{n4}, distributed with
ANTs 2.2.0 \citep[RRID:SCR\_004757]{ants}, and used as T1w-reference
throughout the workflow. The T1w-reference was then skull-stripped with
a \emph{Nipype} implementation of the \texttt{antsBrainExtraction.sh}
workflow (from ANTs), using OASIS30ANTs as target template. Brain tissue
segmentation of cerebrospinal fluid (CSF), white-matter (WM) and
gray-matter (GM) was performed on the brain-extracted T1w using
\texttt{fast} \citep[FSL 5.0.9, RRID:SCR\_002823,][]{fsl_fast}.
Volume-based spatial normalization to two standard spaces
(MNI152NLin6Asym, MNI152NLin2009cAsym) was performed through nonlinear
registration with \texttt{antsRegistration} (ANTs 2.2.0), using
brain-extracted versions of both T1w reference and the T1w template. The
following templates were selected for spatial normalization: \emph{FSL's
MNI ICBM 152 non-linear 6th Generation Asymmetric Average Brain
Stereotaxic Registration Model} {[}\citet{mni152nlin6asym},
RRID:SCR\_002823; TemplateFlow ID: MNI152NLin6Asym{]}, \emph{ICBM 152
Nonlinear Asymmetrical template version 2009c}
{[}\citet{mni152nlin2009casym}, RRID:SCR\_008796; TemplateFlow ID:
MNI152NLin2009cAsym{]},
\item[Functional data preprocessing]
For each of the 14 BOLD runs found per subject (across all tasks and
sessions), the following preprocessing was performed. First, a reference
volume and its skull-stripped version were generated using a custom
methodology of \emph{fMRIPrep}. Susceptibility distortion correction
(SDC) was omitted. The BOLD reference was then co-registered to the T1w
reference using \texttt{flirt} \citep[FSL 5.0.9,][]{flirt} with the
boundary-based registration \citep{bbr} cost-function. Co-registration
was configured with nine degrees of freedom to account for distortions
remaining in the BOLD reference. Head-motion parameters with respect to
the BOLD reference (transformation matrices, and six corresponding
rotation and translation parameters) are estimated before any
spatiotemporal filtering using \texttt{mcflirt} \citep[FSL
5.0.9,][]{mcflirt}. The BOLD time-series (including slice-timing
correction when applied) were resampled onto their original, native
space by applying the transforms to correct for head-motion. These
resampled BOLD time-series will be referred to as \emph{preprocessed
BOLD in original space}, or just \emph{preprocessed BOLD}. The BOLD
time-series were resampled into standard space, generating a
\emph{preprocessed BOLD run in MNI152NLin6Asym space}. First, a
reference volume and its skull-stripped version were generated using a
custom methodology of \emph{fMRIPrep}. Several confounding time-series
were calculated based on the \emph{preprocessed BOLD}: framewise
displacement (FD), DVARS and three region-wise global signals. FD and
DVARS are calculated for each functional run, both using their
implementations in \emph{Nipype} \citep[following the definitions
by][]{power_fd_dvars}. The three global signals are extracted within the
CSF, the WM, and the whole-brain masks. Additionally, a set of
physiological regressors were extracted to allow for component-based
noise correction \citep[\emph{CompCor},][]{compcor}. Principal
components are estimated after high-pass filtering the
\emph{preprocessed BOLD} time-series (using a discrete cosine filter
with 128s cut-off) for the two \emph{CompCor} variants: temporal
(tCompCor) and anatomical (aCompCor). tCompCor components are then
calculated from the top 5\% variable voxels within a mask covering the
subcortical regions. This subcortical mask is obtained by heavily
eroding the brain mask, which ensures it does not include cortical GM
regions. For aCompCor, components are calculated within the intersection
of the aforementioned mask and the union of CSF and WM masks calculated
in T1w space, after their projection to the native space of each
functional run (using the inverse BOLD-to-T1w transformation).
Components are also calculated separately within the WM and CSF masks.
For each CompCor decomposition, the \emph{k} components with the largest
singular values are retained, such that the retained components' time
series are sufficient to explain 50 percent of variance across the
nuisance mask (CSF, WM, combined, or temporal). The remaining components
are dropped from consideration. The head-motion estimates calculated in
the correction step were also placed within the corresponding confounds
file. The confound time series derived from head motion estimates and
global signals were expanded with the inclusion of temporal derivatives
and quadratic terms for each \citep{confounds_satterthwaite_2013}.
Frames that exceeded a threshold of 0.5 mm FD or 1.5 standardised DVARS
were annotated as motion outliers. All resamplings can be performed with
\emph{a single interpolation step} by composing all the pertinent
transformations (i.e.~head-motion transform matrices, susceptibility
distortion correction when available, and co-registrations to anatomical
and output spaces). Gridded (volumetric) resamplings were performed
using \texttt{antsApplyTransforms} (ANTs), configured with Lanczos
interpolation to minimize the smoothing effects of other kernels
\citep{lanczos}. Non-gridded (surface) resamplings were performed using
\texttt{mri\_vol2surf} (FreeSurfer).
\end{description}

Many internal operations of \emph{fMRIPrep} use \emph{Nilearn} 0.6.2 
\citep{abraham_machine_2014}, mostly within the functional
processing workflow. For more details of the pipeline, see
\href{https://fmriprep.readthedocs.io/en/latest/workflows.html}{the
section corresponding to workflows in \emph{fMRIPrep}'s documentation}.

\hypertarget{copyright-waiver}{%
\paragraph{Copyright Waiver}\label{copyright-waiver}}

The above boilerplate text was automatically generated by fMRIPrep with
the express intention that users should copy and paste this text into
their manuscripts \emph{unchanged}. It is released under the
\href{https://creativecommons.org/publicdomain/zero/1.0/}{CC0} license
\clearpage
\newpage
\section{Supplement to results}
\subsection{Do basic statistical differences between HCP and Multi-task data affect transfer performance?}
\label{sec:appendix:results:noise-tests-multi-task}

To better understand whether any basic differences in statistical properties, noise, or preprocessing between the HCP and Multi-task datasets affected the transfer performance of the pre-trained 3D-DeepLight variant, we performed a sequence of additional analyses.

We can immediately rule out basic differences in the temporal distribution of the voxel signals as we detrended and standardized the time series signal of each voxel within each fMRI run (to have a mean of 0 and unit variance; see section \ref{sec:methods:data} of the main text). DeepLight further does not know about the temporal distribution of brain activity as it solely acts on the level of individual fMRI volumes. We therefore next probed the mean and standard deviation of voxel activities within each fMRI volume. We did not find any meaningful differences in the distribution of the volume means and standard deviations between the HCP and Multi-task datasets (see Appendix Fig.\ \ref{fig:appendix:deeplight:transfer:basic-volume-stats}).

We also tested whether other generic differences in noise between the HCP and Multi-task datasets affected transfer performance, by performing a confound correction of the Multi-task fMRI data, in which we regressed out variance related to the six motion correction parameters and three temporal and anatomical noise components resulting from fMRIPrep's CompCor method (for an overview, see Appendix Fig.\ \ref{fig:appendix:deeplight:transfer:finetuning-comparison-multitask-confound-corrected}). Yet, the pre-trained model did not perform better when fine-tuned on the confound-corrected fMRI data than when fine-tuned on the fMRI data that was not confound corrected (for an overview of the training methods, see section \ref{sec:methods:deeplight-training} of the main text). 3D-DeepLight's final decoding accuracy on the confound-corrected data was $43.27\%$, thereby $-2.5\%$ worse than when applied to the uncorrected fMRI data ($t(5)=-4.65, P=0.0056$; Appendix Fig.\ \ref{fig:appendix:deeplight:transfer:finetuning-comparison-multitask-confound-corrected}).

Lastly, we also tested whether the transfer of the pre-trained model to the Multi-task data was affected by the different preprocessing that we applied to both datasets (we preprocessed the Multi-task dataset with fMRIPrep \citep{esteban_fmriprep_2019}, whereas the HCP uses an internal preprocessing pipeline; see section \ref{sec:methods:data} of the main text). To this end, we downloaded the raw fMRI data of another 50 subjects in the HCP working memory task and also preprocessed these with fMRIPrep (for an overview of the preprocessing steps, see Appendix \ref{sec:appendix:preprocessing-fmrirpep-HCP-WM}). Interestingly, the pre-trained 3D-DeepLight variant again exhibited the advantages of transfer learning in this newly preprocessed fMRI dataset, by learning faster and achieving higher decoding accuracies than a model variant that was not pre-trained (see Appendix Fig. \ref{fig:appendix:deeplight:transfer:finetuning-comparison-HCP-WM-fmriprep}). After training on the fMRI data of 20 subjects from this newly preprocessed dataset, the pre-trained model achieved a final decoding accuracy of $72.95\%$ in the fMRI data of the remaining 30 subjects, while the model variant that was not pre-trained achieved a final decoding accuracy of $64.46\%$ (i.e., $-8.49\%$ worse than the pre-trained model, $t(29)=-13.28, P<0.0001$; see Appendix Fig. \ref{fig:appendix:deeplight:transfer:finetuning-comparison-HCP-WM-fmriprep}; for an overview of the training methods, see section \ref{sec:methods:deeplight-training} of the main text).

Overall, we can therefore rule out that the transfer of the pre-trained model to the Multi-task dataset was affected by basic differences in the statistical properties, noise or preprocessing between the HCP and Multi-task datasets.
\begin{figure}[H]
	\centering
 	\includegraphics[width=\linewidth]{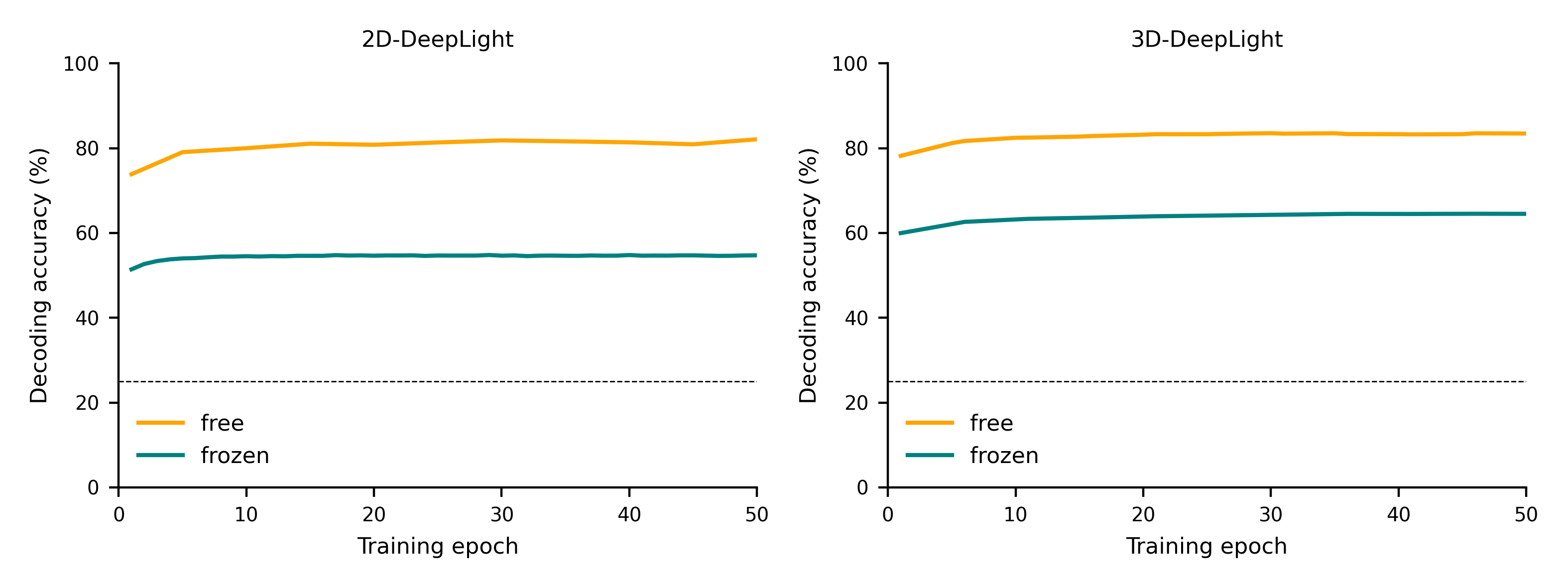}
 	\caption{Comparing two different fine-tuning approaches on the validation data of the HCP working memory task (see section \ref{sec:methods:data:hcp} of the main text). We initialized the weights of two variants of each DeepLight architecture (left: 2D-DeepLight, right: 3D-DeepLight) to the weights of the pre-trained models (all except for the output layer, which now included four instead of 16 neurons; see section \ref{sec:methods:deeplight-architectures} of the main text for an overview of the architectures and Fig.\ \ref{fig:results:pre-training-performance} of the main text for an overview of the pre-trained model performance). We froze the pre-trained weights of one variant of each architecture during fine-tuning (depicted in green), while the other model variant was allowed to train all of its weights during fine-tuning (depicted in yellow) (see section \ref{sec:methods:deeplight-training} of the main text for an overview of the training procedures). Lines indicate decoding accuracy in the validation data as a function of the training epochs. Chance accuracy is indicated by the dashed horizontal line.}
 	\label{fig:appendix:deeplight:transfer:finetuning-approach-comparison}
\end{figure}

\newpage
\begin{figure}[H]
	\centering
 	\includegraphics[width=\linewidth]{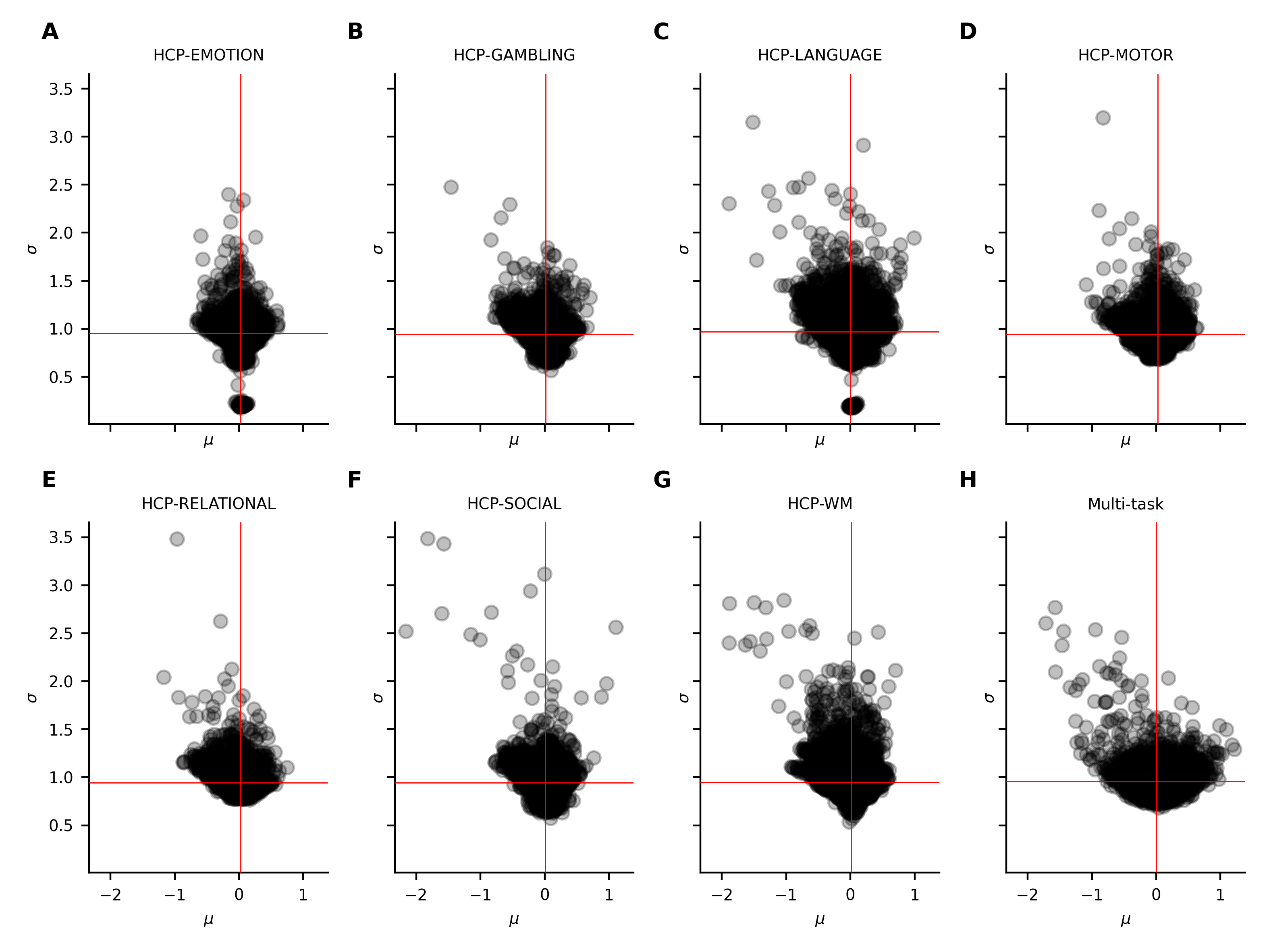}
 	\caption{Mean and standard deviation of voxel activities within each preprocessed fMRI volume in the validation datasets of the HCP experimental tasks (A-G) and Multi-task data (H) (for an overview of the datasets, see section \ref{sec:methods:data} of the main text). Scatter points indicate individual fMRI volumes. Red lines indicate the mean over volumes.}
 	\label{fig:appendix:deeplight:transfer:basic-volume-stats}
\end{figure}

\newpage
\begin{figure}[H]
	\centering
 	\includegraphics[width=0.6\linewidth]{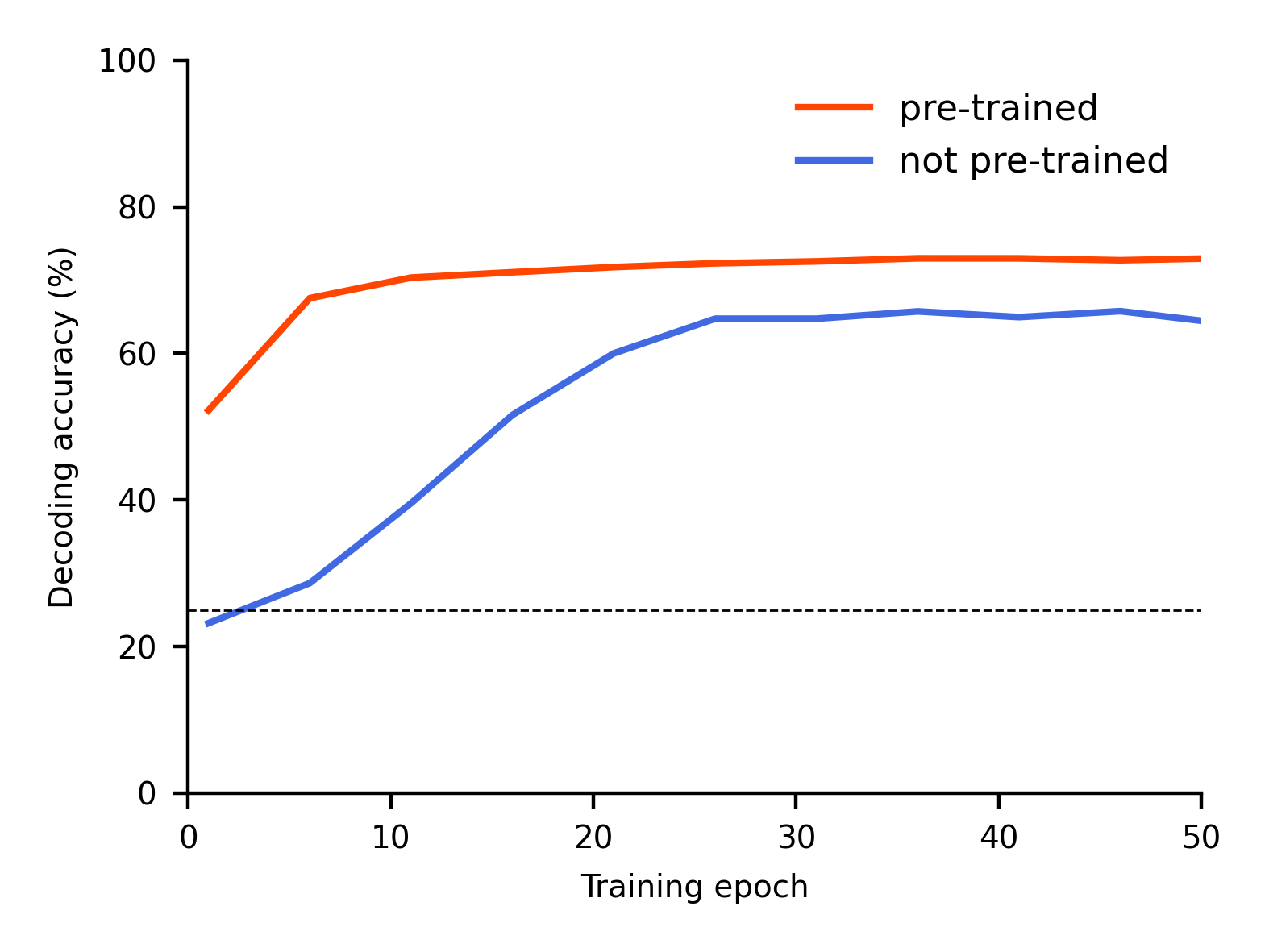}
 	\caption{Training decoding accuracy for a pre-trained (red) and not pre-trained (blue) 3D-DeepLight variant in the validation data of the HCP working memory task that was preprocessed with fMRIPrep (see Appendix \ref{sec:appendix:results:noise-tests-multi-task}; see section \ref{sec:methods:deeplight-training} of the main text for an overview of the training procedures). An epoch was defined as an entire iteration over the training dataset. Lines indicate decoding accuracy. Chance accuracy is indicated by the dashed horizontal line.}
 	\label{fig:appendix:deeplight:transfer:finetuning-comparison-HCP-WM-fmriprep}
\end{figure}

\newpage
\begin{figure}[H]
	\centering
 	\includegraphics[width=0.6\linewidth]{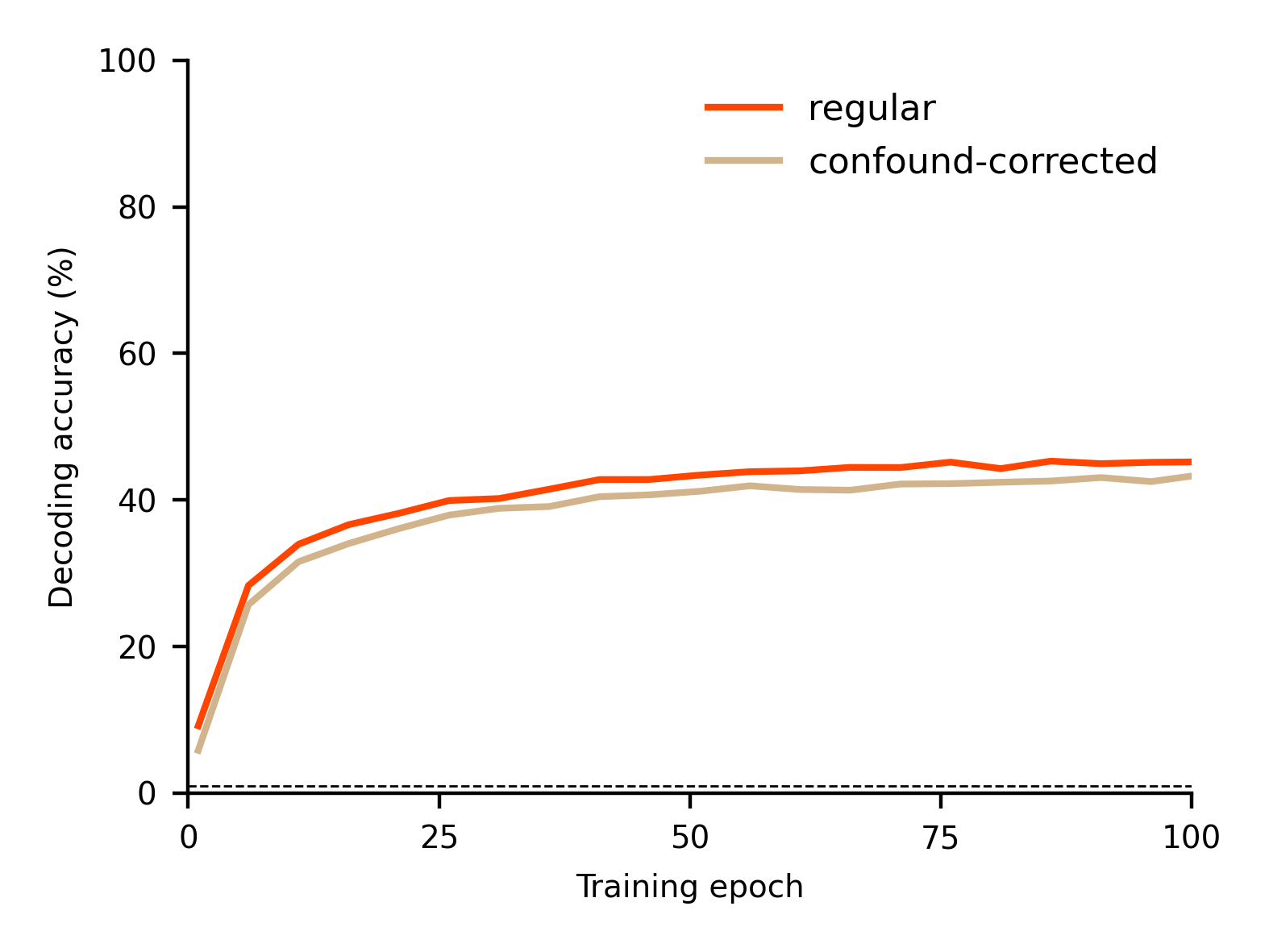}
 	\caption{Training decoding accuracy for the pre-trained 3D-DeepLight variant in two conditions: when it is fine-tuned on the regular fMRI data of the Multi-task dataset (red) or on a version that is corrected for basic noise confounds (tan). Specifically, we corrected the Mutli-task data for any variance resulting from the six parameters of basic motion correction, as well as the three temporal and anatomical noise components with the largest singular values resulting from fMRIPrep's CompCor method (for details on this method, see \citet{compcor}), by regressing their variance out of the time-series signal of each voxel \citep[as implemented in Nilearns "signal.clean" function; ][]{abraham_machine_2014}. See section \ref{sec:methods:deeplight-training} of the main text for an overview of the training procedures. An epoch was defined as an entire iteration over the training dataset. Lines indicate decoding accuracy. Chance accuracy is indicated by the dashed horizontal line.}
 	\label{fig:appendix:deeplight:transfer:finetuning-comparison-multitask-confound-corrected}
\end{figure}

\newpage
\begin{figure}[H]
	\centering
 	\includegraphics[width=\linewidth]{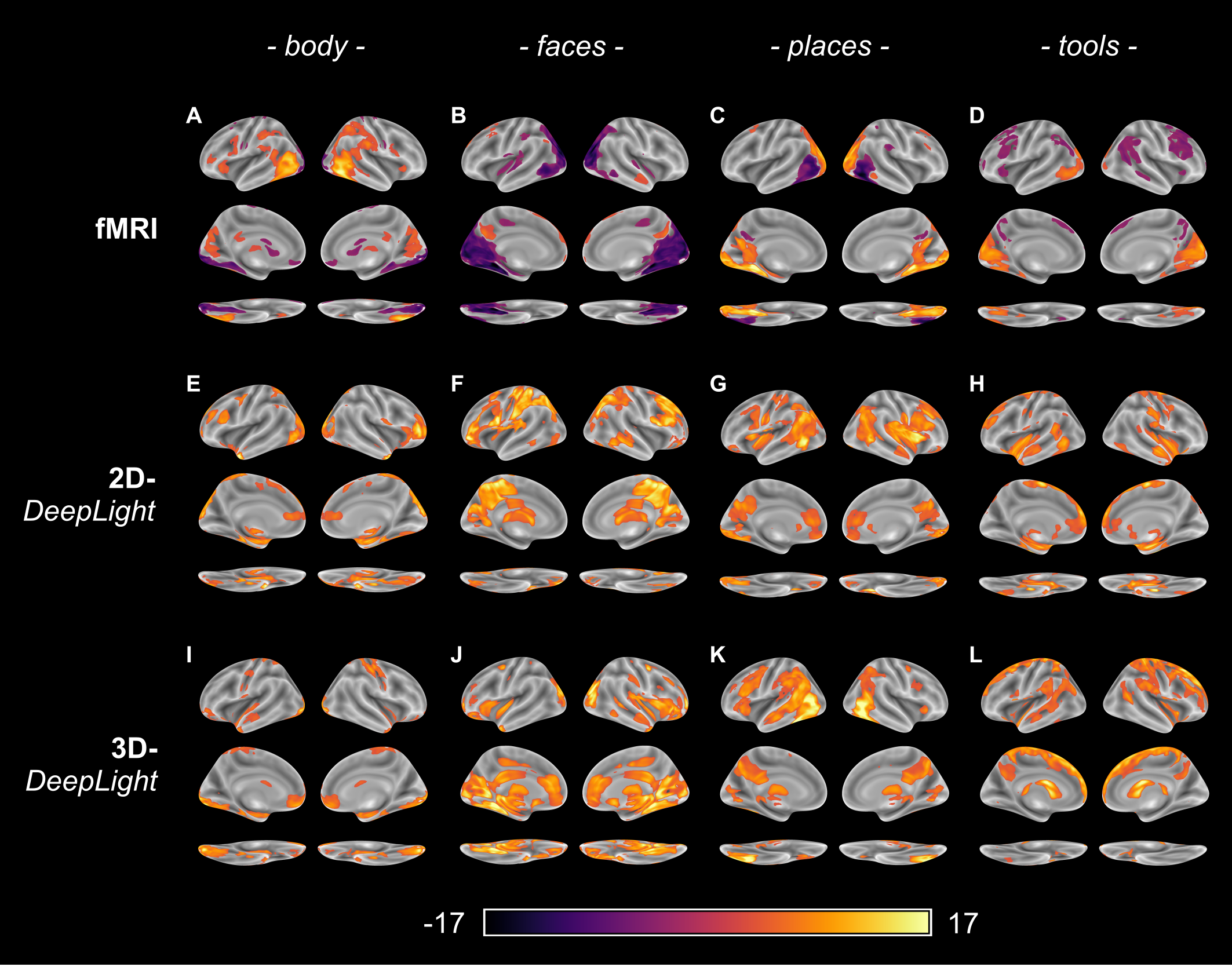}
 	\caption{Learned mappings between brain activity and cognitive states of the pre-trained DeepLight variants that were fine-tuned on the full training dataset of the HCP-WM experimental task (see section \ref{sec:results:fine-tuning} of the main text). A-D: We first computed a standard two-stage GLM analysis \citep{holmes_generalisability_1998} of the fMRI data of the 50 subjects in the validation dataset of this task. E-L: We then also interpreted the decoding decisions of the 2D- (E-H) and 3D-DeepLight (I-L) variants for the same data. To identify the brain regions that each DeepLight variant associates most strongly with a cognitive state, we computed a similar two-stage GLM analysis of the resulting relevance data (restricting the resulting z-scores to only positive values). All GLM analyses were performed on parcellated brain data by the use of the dictionaries for functional modes (DiFuMo) atlas with 256 brain networks \citep{dadi_fine-grain_2020} and computed separately for each experimental task by contrasting each cognitive state of the task against all other states of that task (for details on the GLM analysis, see Appendix \ref{sec:appendix:methods:glm-analysis-details}). All brain maps are thresholded at a false-discovery rate of 0.001 and projected onto the inflated cortical surface of the FsAverage template \citep{fischl_freesurfer_2012}. Brighter yellow values indicate larger z-scores.}
 	\label{fig:appendix:deeplight:fine-tuning-brainmaps}
\end{figure}
\clearpage
\newpage

\end{document}